\documentstyle[aps]{revtex}
\begin{document}
\draft
\title{Variational Master Field for Large-N Interacting Matrix Models -- 
Free Random Variables on Trial }
\author{M. Engelhardt$^{1,2,}$\thanks{Address from 1.10.96: Institut f\"ur
theoretische Physik, Universit\"at T\"ubingen, Auf der Morgenstelle 14,
72076 T\"ubingen, Germany.}$^{,}$\thanks{Supported in part by MINERVA
and by the Bundesministerium f\"ur Bildung und Forschung, Germany.}
and S. Levit$^{1}$ }
\address{$^{1}$Department of Condensed Matter Physics \\
Weizmann Institute of Science \\ Rehovot 76100, Israel \\
$^{2}$Institut f\"ur theoretische Physik III \\
Universit\"at Erlangen \\ Staudtstr.7, 91058 Erlangen, Germany}
\date{}
\maketitle

\begin{abstract}
Matrices are said to behave as free non-commuting random
variables if the action which governs their dynamics constrains only their
eigenvalues, i.e. depends on traces of powers of individual matrices.
The authors use recently developed mathematical techniques in combination 
with a standard variational principle to formulate a new variational
approach for matrix models. Approximate variational solutions of
interacting large-N matrix models are found using the free random matrices 
as the variational space. Several classes of classical and quantum mechanical
matrix models with different types of interactions are considered and 
the variational solutions compared with exact Monte Carlo and analytical
results. Impressive agreement is found in a majority of cases.
\end{abstract}


\section{Introduction} 
The main motivation behind this work lay in
developing approximate analytical tools for dealing with the long range 
physics of
Quantum Chromodynamics. The large-$N_c$ approximation is an attractive
candidate for this project. Initiated by 't Hooft, Ref. \cite{Hoo1}, 
it was immediately applied to solve large-$N_c $
1+1 dimensional QCD, Refs. \cite{Hoo2}, \cite{Col},
\cite{Cal}, \cite{Ein} and led to extremely fruitful phenomenological insights 
and models, Refs.
\cite{Col}, \cite{Wit}, \cite{Adk}. The factorization property of 
correlators of gauge
invariant operators suggested the idea of the master field -- the gauge field
configuration which dominates the path integral in the
large-$N_c$ limit, cf. Refs.
\cite{Col}, \cite{Wit}, \cite{Mak}. Knowledge of the master field should 
allow to
calculate any gauge invariant observable as if it were a 
classical non-fluctuating object.

A concrete example of a master field was provided by the exact solution
of a model which described the dynamics (classical as well as 
quantum mechanical) of
a single hermitian $N\times N$ matrix, Ref. \cite{Bre1}. It was found 
that in the
large-$N$ limit the Boltzmann integral (in the classical case) 
and the ground state properties
(in the quantum case) of this model were completely determined by an
ensemble of matrices with a frozen distribution of eigenvalues given by a
solution of a certain integral equation of the mean field type.  The 
``angular'' orientations of the matrices in this ensemble, i.e.
the unitary matrices $W$ in the
decomposition $M = WmW^{+}$ with $m= {\rm diag}(m_1, m_2, . . . , m_N)$
were completely free, i.e. distributed with equal probability according to the
unbiased $U(N)$ group theoretical Haar measure. This work and, in its sequel,
Ref. \cite{Itz} gave rise to an intense interest in solutions of large-$N$
matrix models and made contact with the study of such models in other fields
of physics, most notably nuclear physics, cf. Ref. \cite{Meh} and, more
recently, two-dimensional quantum gravity, cf. Ref.~\cite{Bre2} and quantum 
chaos, cf. Ref. \cite{Alt}. The
general interest in large-$N$ matrix models has been behind a steady progress
in understanding solutions of various versions and generalizations of such
models as reported e.g. in Refs. \cite{Amb}, \cite{Bre3}, \cite{Bre4}, 
\cite{Kaz}, \cite{Kazm}.

A new direction in this activity opened up when the physics
community became acquainted with recent advances in the
mathematical literature, associated with
the work of Voiculescu, Ref. \cite{Voi}, on non-commutative probability
theory, Ref. \cite{Sin}.  The statistical mechanics (classical or 
quantum mechanical) of matrix
(non-commuting) variables or (gauge) fields is an example of probability
theory of non-commutative variables. A concept which turned out to be
especially useful for the study of large-$N$ matrix and gauge theories, 
cf. Refs. \cite{Gop} and
\cite{Dou}, was the introduction in Ref. \cite{Voi} of the so-called 
{\em free random variables}. The
simplest examples of such variables occur in independent large-$N$ matrix
models, i.e. in the classical statistical mechanics of several 
$N\times N, \;\; N\to\infty $
hermitian matrices $M_i,\, i=1, \ldots ,D$,
with the probability distribution (the Boltzmann factor) 
given as $const \cdot \exp(-N^2 \sum_i V_i (M_i))$ with $V_i (M_i ) =
(1/N) \mbox{Tr} \, P_i (M_i )$ and $P_i $ 
polynomial functions. 
Typical ``observables'' in such a model are correlators 
\begin{equation}
\frac{1}{N}\langle \mbox{Tr} \, (M_{i_1} \ldots
M_{i_k})\rangle
\equiv \int \, \frac{1}{N} \, \mbox{Tr} \,
(M_{i_1} \ldots M_{i_k})
\exp\{\sum_{i=1}^D [F_i - N^2 V_i(M_i)]\}\prod_{k=1}^{D} D\mu[M_k] \; , 
\end{equation}
with the integration measure
\begin{equation}
D\mu[M_k] \equiv \prod_{\gamma=1}^N d(M_k)_{\gamma\gamma}
\prod_{\gamma > \nu =1}^N d \mbox{Re} (M_k)_{\gamma\nu}
\;d \mbox{Im} ( M_k)_{\gamma\nu} \; , \label{intmes}
\end{equation}
and $F_i = -\ln \int D\mu[M_i]
\exp[-N^2 V_i(M_i)] \; . $ 
Were the $M_i $ regular commuting variables, one would call them
independent since $\langle M_{i_1} \ldots M_{i_k}\rangle =
\langle M_{i_1}\rangle \ldots \langle M_{i_k}\rangle $ for any selection
$i_1 ,\ldots , i_k$. For the matrix model this is obviously not the case.
The probability distribution depends only upon the eigenvalues
$m_i$ of the
$M_i$. The expectation of products of the $M_i$ on the other hand involves
also the ``angular'' variables
$W_i$ in the decomposition $M_i = W_i m_i W_i^{+}$.  These variables
however are ``free'', meaning that they are weighted with the unbiased
Haar $U(N)$ measure. The integrals over the
$W_i$ can therefore be evaluated in a universal manner independent of what
the potentials $V_i$  are.  The remaining integrals over the $m_i$ can then
be evaluated in the limit $N\to\infty$ by a saddle point approximation 
as in Ref.
\cite{Bre1}. The remarkable fact is that this
procedure can be formulated in terms of
very general rules relating any correlator
of independent matrices to a
linear combination of products of various individual moments of these 
matrices, cf. \cite{Voi}. 
These rules will be reviewed
in Section \ref{tracfr} below. Any non-commutative
variables distributed in such a manner that they satisfy these rules are 
termed
``free random variables''. Refs. \cite{Gop} and \cite{Dou} provide a 
review of the
techniques which were developed to deal with such things as sums and 
products of
free random variables, to define a certain operator algebra in a suitably 
defined Fock space, and further applications, cf. also \cite{Zee}, 
\cite{Are}, \cite{Nsp}.

It has been possible to prove that a master field for any matrix model
can be constructed on the basis of the operator algebra developed for free
models, Ref. \cite{Gop}, \cite{Dou}.
However, so far no dynamical principle has been 
advanced in order to carry out this construction explicitely for interacting   
matrix and field theories. This paper presents an attempt to make progress 
in this
direction by using a variational method based on the inequality 
$\langle e^{-S}
\rangle \ge e^{-\langle S\rangle } $.
This approach is fairly standard in 
statistical physics.  
The new element will be to use the free matrix models as the trial variational
space. The
details of the development are outlined in Section \ref{secone} below.
Given a certain set of variables in terms of which the action is formulated,
one can give a complete solution of the variational problem in the space
of free matrix models by defining
what will be called the free partner of the exact action of the model.
Subsequently one observes
however that the available variational freedom is even wider since one 
can first transform to a new set of variables, $M_i\to M_i({B_j})$.
In the new $B_i$ it is again
possible to define the free partner of the action and give the general
variational solution. Thus the ``art part'' of the proposed
variational method is the choice of the optimal $B_i$ or in 
other words of the
optimal parameters of the transformation $M_i\to M_i({B_j})$. 

Unfortunately it is hard to give estimates of the accuracy of 
this variational approximation. As usual, one must gain experience
from examples and comparison with exact solutions. In Section
\ref{secexa}, a number of such examples are considered starting with
classical interacting matrices and then going to
quantum models of first one and then two interacting quantum
matrices. Only linear transformations
$M_i=\sum_j  c_{ij}B_j $ are considered and the variational results
using free random variables are compared with exact Monte Carlo
calculations and, where available, with analytic results. In almost
all models considered, e.g. in the models of 
two interacting matrices with the action 
$$
S= \frac{1}{N} \mbox{Tr} \, (M_1^4 + M_2^4 + g
M_1^2 M_2^2)
$$ 
and various values of $g$, of many interacting matrices with
$$
S=\frac{1}{D-1}
\sum_{i\ne j}^D \frac{1}{N} \mbox{Tr} \, M_i^2 M_j^2
\;\;\; {\rm  and} \;\;\; 
S = \frac{1}{D-1} \sum_{i\ne j}^D \frac{1}{N} \mbox{Tr} \, M_i^4 M_j^4
$$ 
and in the quantum models of one and two interacting matrices,
impressively good agreement is found
with exact results. Some models are also presented for which the method 
(with the additional limitation 
to the linear $M\to B$ transformations) failed completely.
Presumably nonlinear
transformations could improve the agreement but the authors have not
investigated this.   

\section{Variational Method for Matrix Models} \label{secone}

In this paper, matrix models described by partition functions 
of the type
\begin{equation}
Z\equiv e^{-F} = \int \exp (-N^2 S[M_1,. . .,M_D])
\prod_{k=1}^D D\mu[M_k]
\label{matint}
\end{equation}
are considered, with $M_k$, $k = 1,\ldots ,D$, 
representing hermitian $N \times N$
dimensional matrices, in the limit $N\rightarrow \infty $. A
factor $N^2 $ has been pulled out in the exponent and in return,
all matrix traces appearing in actions $S$ will be accompanied
by a factor $1/N$.
Note that the term ``action'' is used very loosely here to denote
essentially any weight appearing in the exponent of the Boltzmann
factor in a partition function. Only actions
$S[M_1,\ldots ,M_D]$ which are invariant under the global unitary 
``rotations''
$M_i\to U M_i U^{+}$ will be considered. One can further distinguish 
between ``classical'' and ``quantum''
matrix models. By classical one means the models for which the 
above integral
is just the ordinary integral with the integration measure given by
(\ref{intmes}) whereas in quantum models this is a path
integral with $M_i$ depending on (imaginary) time $\tau$, $M_i(\tau)$, and
the measure
\begin{equation}
{\cal N}\prod_{0<\tau<\beta} D\mu[M_k(\tau)] =
{\cal N}\prod_{0<\tau<\beta}\prod_{\gamma=1}^N
d(M_k(\tau))_{\gamma\gamma}
\prod_{\gamma > \nu =1}^N d \mbox{Re} (M_k(\tau))_{\gamma\nu} \;d \mbox{Im}
(M_k(\tau))_{\gamma \nu} \; . 
\end{equation}
Here $\beta$ is the inverse temperature and
$\cal N$ is a normalization factor which in principle depends on $\Delta \tau$, 
but will not be important in the following.

In both the classical and quantum cases one deals with integrals over many
matrix variables and the real distinction will lie in the form of the action. 
In the classical
case the action (the classical potential) is a {\em real function} 
usually represented
as a sum of monomials in powers of the $M_i$,
\begin{equation}
S_{\rm cl} \equiv V(M_1, \ldots ,M_D) = \frac{1}{N} \,
\sum_{i_1,i_2 ,\ldots ,i_D} \sum_{m_1,m_2 ,\ldots ,m_D} C_{m_1 m_2 \ldots
m_D}^{i_1,i_2,\ldots ,i_D} 
\mbox{Tr} \, \left(M_{i_1}^{m_1}M_{i_2}^{m_2} \ldots M_{i_D}^{m_D}\right)\; ,
\label{clact}
\end{equation}
with real coefficients
$C_{m_1 m_2 \ldots m_D}^{i_1,i_2, \ldots ,i_D}$. In
the quantum case the action will be a sum of kinetic and potential terms
\begin{equation}
S_Q =
\frac{1}{2\Delta
\tau}\sum_{0<\tau<\beta}\;\;\sum_{i=1}^D 
\frac{1}{N} \mbox{Tr} \, \left[M_i(\tau +\Delta \tau) -
M_i(\tau)\right]^2 + \sum_{0<\tau<\beta} \Delta\tau V(M_1(\tau), \ldots ,
M_D(\tau)) \; , 
\end{equation}
with $V$ as in (\ref{clact}).
    
\subsection{The Variational Principle}

As was reviewed in the Introduction, only a limited number of matrix models are
amenable to analytical solution so that various methods of approximation 
are called for.
Here, the powers of the variational method, combined with the large-$N$
saddle point evaluation, will be explored.
As is common with statistical integrals,
the variational approximation will be developed
on the basis of the inequality, cf. Ref. \cite{Fey}, 
\begin{equation}
\Big< e^{- x}\Big> \ge e^{-\left<x\right>} \; ,
\label{ineq}
\end{equation}
which expresses the convexity property of the exponential
function -- the average value of the exponential of a random variable $x$ is
greater than or equal to the exponential of the average provided $x$ is real 
and the weights used in the averaging are positive. 

If one can find an action $S_0[M]$ which is simple enough to allow the 
calculation of integrals like 
$$
\int \prod_k  D\mu[M_k ]e^{-N^2 S_0[M]}\;\;\;\; {\rm and
}\;\;\;\;  \int \prod_k D\mu[M_k] K(M_1, \ldots , M_D)e^{-N^2 S_0[M]} \; 
$$ 
for some simple
functions $K$ then one can approximate the integral (\ref{matint}) as follows. 
One can represent
\begin{equation}
e^{-F} = \int \prod_k D\mu[M_k]e^{-N^2 S[M]}
= \int \prod_k D\mu[M_k] e^{-N^2 (S[M] - S_0[M])} e^{-N^2 S_0[M]} 
= e^{-F_0} \Big< e^{-N^2 (S - S_0)} \Big>_{S_0 } \; .
\end{equation}
where  $\langle O\rangle_{S}$ means
$\int O\exp(F - N^2 S) D\mu(M)$. Using the inequality (\ref{ineq}) one obtains 
\begin{equation}
e^{-F} = e^{-F_0}\Big< e^{-N^2 (S - S_0)} \Big>_{S_0}
\ge e^{-[F_0 + N^2 \big<S - S_0 \big>_{S_0} ] } \;,
\end{equation}
or 
\begin{equation}
F\le F_0 + N^2
\big<S - S_0 \big>_{S_0} \; . \label{minp} 
\end{equation}
If $S_0$ depends on adjustable
parameters, one can find their optimal values by minimizing the right hand
side of the above inequality. The resulting
$S_0$ can then be adopted as an approximation for $S$ with which one should
calculate the relevant correlators. 
As always with variational methods, it is hard to
assess the accuracy of the approximation. Here, it will be tested in
different classes
of matrix models, using the so-called free matrix models (cf. below) as the
variational space, by comparing (where available) with exact analytical
results or with Monte Carlo simulations. 
 
\subsection{Free Matrix Models} \label{tracfr}

Freeness is a concept recently developed in the mathematical literature   
associated with noncommuting random variables
\cite{Voi}. While freeness is abstractly defined without reference to a 
specific
realization of the random variables, in the context discussed here, namely in 
relation to large-$N$ hermitian multi-matrix models, it essentially means  
that the weight
$\exp(F-N^2 S)$ according to which  the matrices are distributed contains no 
bias with respect to the relative $SU(N\rightarrow \infty )$ ``orientations''
of 
the matrices. In terms of the decomposition $M_k  = W_k m_k W_k^{+}$ with
diagonal $m_k$ and unitary $W_k$ this means that the action $S$ in the free
models depends only on the $m_k$. Consequently all the angular integrals in any 
given correlation function are weighted only with the group theoretical $U(N)$
Haar measure and give universal results independent of what the action is. The
remaining integrals over the $m_k$ in the large-$N$ limit are controlled by 
the saddle
point of the effective action which includes the contribution from the
Vandermonde determinants for each $k$, cf. examples below. 
As a consequence of these  
properties all correlation functions for free multi--matrix models are 
systematically accessible in terms of the {\em moments of individual matrices}
through the recently developed techniques, as will be elaborated
upon below,
cf. also Refs. \cite{Voi}, \cite{Gop}, \cite{Dou}.

A family of noncommuting random variables $M_i$ is called free if and only if 
\begin{equation}
\langle f_1 (M_{i_1 } ) f_2 (M_{i_2 } ) \ldots f_n (M_{i_n } ) \rangle =0 \
\mbox{whenever} \ \langle f_j (M_{i_j } ) \rangle = 0 \ \mbox{for all} \ j 
\end{equation} 
where $i_j \neq i_{j+1} $ (and $i_{n+1} $ is identified with $i_1 $ ).
Note that the bracket notation introduced
here following \cite{Voi} means taking the expectation value of the
trace of the expression inside, divided by $N$.
In order to evaluate a general
correlation function $\langle p_1 (M_{i_1 } ) p_2 (M_{i_2 } ) \ldots
p_n (M_{i_n } ) \rangle $, one can use 
\begin{equation}
\langle (p_1 (M_{i_1 }  ) - \langle p_1 (M_{i_1  } ) \rangle ) \; 
(p_2 (M_{i_2 } )  -
\langle p_2  (M_{i_2 } ) \rangle ) \; \ldots \;
(p_n (M_{i_n } ) - \langle p_n (M_{i_n } ) \rangle ) \rangle = 0 
\label{algo}
\end{equation}
and multiply out the terms in the expectation value. This
recursively determines correlation functions of a certain order in terms of
lower-order correlation functions until one arrives at an expression containing
only {\em the moments of the individual matrices}. For two free matrices
$M_1 $ and $M_2 $ this algorithm yields, e.g., 
\begin{eqnarray}
\langle p_1 (M_1 )  p_2 (M_2 ) \rangle  &=& \langle p_1 (M_1 ) \rangle
\, \langle p_2 (M_2  ) \rangle \\ \langle M_1  M_2 M_1 M_2 \rangle &=&
\langle M_1^2   \rangle \, \langle   M_2  \rangle^{2} + \langle  M_2^2
\rangle \, \langle M_1  \rangle^{2}    - \langle M_1  \rangle^{2}   \,
\langle M_2 \rangle^{2}
\label{m1m2e} 
\end{eqnarray}

Building on this information, it has been possible, e.g., to determine,
\cite{Voi}, how the eigenvalue distributions of matrices which are free with
respect to one another are additively and multiplicatively convoluted. 

Freeness in matrix models is a slightly more general concept than what is
usually meant by the term ``independent matrix model'', in which the potential
governing the distribution of the matrices is a sum of terms for each
individual matrix. Independent matrix models are certainly free; however, 
so is e.g. a two-matrix model with potential 
\begin{equation} V(M_1 ,M_2  )  = \frac{1}{N} 
\mbox{Tr} \, M_1^2   + \frac{1}{N} \mbox{Tr}  \,
M_2^2 +
\frac{1}{N^2 } \mbox{Tr} \, M_1^2 \, 
\mbox{Tr} \, M_2^2 \label{freex} \; . 
\end{equation} 
More generally, the potential of a free matrix model can be any function 
of traces over
functions of the individual matrices\footnote{To the authors' knowledge,    
this type of model was first considered in
\cite{Wadia}.}. Such more general free matrix models are almost as easy 
to solve as 
the independent ones. This is best illustrated by an example. 
Consider the model 
described by (\ref{freex}). After the transformation of variables $M_i  = W_i
m_i W_i^{+}$, with diagonal
$m_i$ and unitary $W_i$, the potential $V(M_1, M_2)$ depends only on the
$m_i$. Moreover, the transformation introduces the Vandermonde
Jacobians which, when exponentiated, modify the potential into 
\begin{eqnarray}
N^2 V_{eff}  &=& \sum_{i=1,2} \;\;
\sum_{\alpha\ne\beta}^N\;\ln|m_{i,\alpha }  - m_{i,\beta }|
+  N^2 V(m_1,m_2) = \nonumber \hspace{5.0cm} \\
&=& N^2\left[\sum_{i=1,2}\left(\int_{-\infty}^{\infty}
\rho_i(\mu)\;\ln|\mu-\mu'|\;\rho_i(\mu')\; d\mu \;d\mu'
+\int_{-\infty}^{\infty}\rho_i(\mu)\mu^2\right) +
\prod_{i=1,2}\int_{-\infty}^{\infty}\rho_i(\mu)\mu^2 d\mu\right] \; .
\nonumber
\end{eqnarray}
In the large-$N$ limit, the distributions $\rho_i(\mu)$ of the eigenvalues
are ``frozen'' at a saddle point, so that the moments of the matrices 
have definite values
\begin{equation}
\frac{1}{N} \mbox{Tr}  \,  M_1^2 = x_1  \ \ \ \ \ \ \ \ \frac{1}{N}
\mbox{Tr} \, M_2^2 = x_2 
\label{freeea}
\end{equation} 
This means that the saddle point eigenvalue distributions
$\rho_i(\mu)$ can be solved for as if they were controlled by the potentials 
\begin{equation} 
V(M_1 ) = (x_2 +1) \frac{1}{N} \mbox{Tr} \, M_1^2 \ \ \ \ \ \ \ \  V(M_2 ) =
(x_1 +1) \frac{1}{N} \mbox{Tr} \, M_2^2 \; . 
\label{freeeb}
\end{equation} 
For such quadratic potentials, one obtains the well-known semicircular
distributions, cf. Ref. \cite{Meh},\cite{Bre1}
\begin{equation}
\rho_1 (\mu ) = \frac{x_2  +1}{\pi } \sqrt{\frac{2}{x_2 +1} -\mu^{2} }
\ \ \ \ \ \ \ \ \rho_2 (\mu ) = \frac{x_1 +1}{\pi } 
\sqrt{\frac{2}{x_1 +1} -\mu^{2} }\; .
\end{equation} 
The constants  $x_1 $ and $x_2 $ can now be determined by the
self-consistency conditions 
\begin{equation}
\left. 
\begin{array}{c}
x_1 = \langle M_1^2 \rangle = \int d\mu \, \mu^{2} \rho_1(\mu ) \\
x_2 = \langle M_2^2 \rangle = \int d\mu \, \mu^{2} \rho_2 (\mu ) 
\end{array} \right\}
\Rightarrow x_1 = x_2 = \frac{\sqrt{3} -1}{2}\; . 
\end{equation} 
Including such self-consistency conditions on selected moments of
the matrices is the only additional step needed compared with solving
independent matrix models. Knowledge of the eigenvalue distributions
$\rho_i (\mu)$ is sufficient to calculate any correlation function since 
the latter can be reduced using Eq. (\ref{algo}) to a sum of products of
individual moments, i.e. to the integrals $\langle M_i^k \rangle =
\int\mu^k\rho_i(\mu) d\mu $. 

It was argued e.g. in Refs. \cite{Gop},\cite{Dou} that a free
behaviour of the matrix variables representing physical degrees of
freedom may capture correctly the main features of the dynamics in
the large-$N$ limit of nonabelian gauge theories. This hope and the wealth
of available analytical techniques presented above render the space of free
matrix models a good candidate for variational calculations. The main
idea of the present approach is to choose the optimal action $S_0$ in
the variational principle (\ref{minp}) from among all actions which
leave the matrices in terms of which $S$ is given,
free with respect to one another. 

\subsection{General Variational Solution in the Space of Free Matrix
Models}

In making a variational approximation based on free matrix models one
must first decide which combinations of the original 
matrix variables $M_i$ will be assumed to be free, i.e. in which variables
to formulate the exact action $S$. One should realize that
e.g. free $M_1 $, $M_2 $ do not imply free $M_1 +M_2 $, $M_1 -M_2 $ nor
vice versa (more about this below).
The second step is to find the best trial action $S_0$ which is
free in the variables one has settled for.

In this subsection, the solution of the second step is addressed.
In other words, consider $S_0 $ to be {\em free in terms of the
original} $M_i$. Under this limitation, it is possible to give a 
{\em general recipe}
for finding the trial $S_0$ which minimizes the right hand side of 
(\ref{minp}). 
For this it is useful to introduce the following algorithm. Given a trace of
any function of matrices $M_i$, $P=(1/N) \mbox{Tr} \,p(M_1, \ldots , M_D)$,
write down its expectation 
value in terms of the moments of the individual matrices under the assumption
that the matrices are free with respect to one another. 
This is easily accomplished
for any multinomial expression using Eq. (\ref{algo}). Now rewrite the
resulting expression with
$\langle M_i^j \rangle$ replaced by Tr$\, M_i^j /N$. This defines what
will be called
the ``free partner'' $P_{f}$ of $P$ (the ``liberated'' $P$, so to speak).
For instance, the free partner of
$P=(1/N) \mbox{Tr} \,(M_1M_2M_1M_2)$ is 
$$
P_{f} =\frac{1}{N^3 } \mbox{Tr} \, M_1^2 \, (\mbox{Tr} \, M_2)^2
+\frac{1}{N^3 } (\mbox{Tr} \, M_1)^2 \, \mbox{Tr} \, M_2^2 
- \frac{1}{N^4 } (\mbox{Tr} \, M_1)^2 \, (\mbox{Tr} \, M_2)^2 \; ,
$$  
cf. Eq. (\ref{m1m2e}).
Evidently, the original function and its free partner have the
same expectation values when evaluated in any free matrix model for the  
$M_i $ (note that the expectation values of the products of traces
appearing in the free partner factorize at large $N$).
In general, the two expectation values will of course differ. 
Furthermore, if the free partner $S_{f}$ of any action
$S$ is used as a trial action, it will generate a free matrix model; hence the
terminology. 

With the help of the above definitions, one can give the general solution of
the minimization principle (\ref{minp}) in the space of matrix models which 
are free
in terms of the variables $M_i $. The crucial observation is that, when  
$S_0 $ is to
be chosen from among all free actions, the original action $S$ 
and its free partner
$S_{f} $ lead to the same minimization problem, i.e. identical right hand 
sides of Eq.
(\ref{minp}), since $\langle S\rangle_{S_0 } =\langle S_{f}
\rangle_{S_0 } $ by construction for any free $S_0$, as mentioned above.
However, the minimization
problem for $S_{f} $ is trivial to solve: Since $S_{f} $ itself describes a 
free matrix model, the best approximation to $S_{f} $ in the space
of free actions is $S_0 =S_{f} $ itself. Since the approximation to $S$
is governed by the same minimization problem,
$S_0 =S_{f} $ is at the same time the best free approximation to $S$. 

This result provides a general solution of the minimization problem in
the space of matrix models which are free {\em in terms of the original
variables} $M_i $.

Note that the solution $S_f $ of the minimization problem defines a type
of mean field approximation to the exact action $S$. Interaction terms
in $S$ which are sensitive to angular correlations between the different
matrices are replaced by terms in which each matrix couples only to 
selected moments of the other matrices, i.e. to some
mean properties independent of the angular orientations. This is
entirely analogous to the standard development of the Hartree or
Hartree-Fock mean field theories from a variational principle using the
space of (properly symmetrized or antisymmetrized) product states.
There, in second-quantized language, the combinations in which the mean
field enters the single-particle Hamiltonian are determined by Wick's
theorem; here, in complete analogy, the combinations in which the moments
enter the free partner are determined by the free random variable axioms
as encoded in Eq. (\ref{algo}). In fact, (\ref{algo}) is nothing but
Wick's theorem generalized from the usual bosonic and fermionic cases
to the case of objects obeying the Cuntz algebra (this is the algebra
obeyed by the Fock space operator representation of free variables,
cf. Ref. \cite{Gop}).

\subsection{Variable Transformations}

The idea of the present variational approach was stated as choosing the optimal
action $S_0 $ from among all matrix model actions which leave the matrices 
in terms of which $S$ is given, free
with respect to one another. In this there exists an additional freedom 
of choice
which will now be exploited, namely the choice of matrix 
variables in terms
of which $S$ is given, i.e. in terms of which one develops the mean
field approximation by constructing the free partner $S_f $.
Up to now, only a fixed set of variables was considered -- the
original $M_i $. However, if one rewrites the action $S$ in terms of 
other variables, 
\begin{equation} 
S(\{ M_i \} ) = S( \{ M_i ( \{ B_j \} ) \} ) = \tilde{S} ( \{ B_j \} ) 
\end{equation} 
and again looks for the optimal free action, now in terms of the new
variables $B_j $, approximating $\tilde{S} $, one may find a better 
approximation
than the one found using the variables $M_i $. It should be emphasized that free
$B_j $ in general do not imply free $M_i $ nor vice versa. Therefore, 
the accessible
variational space using free matrix models is in fact much larger than was  
apparent in the last section. By allowing different sets of 
variables in which to
formulate the problem to be approximated, one can even include into the
variation models in which the original matrices $M_i $ are not free with 
respect to
one another. To formalize this idea, it is necessary to 
reexamine the derivation of
the variational principle (\ref{minp}). Carrying out a variable transformation
$M_i \rightarrow M_i ( \{  B_j \} )$, one has
\begin{equation} 
e^{-F} = \int  \prod_j D\mu[B_j] e^{-N^2 \tilde{S} [B_j]  + \ln J [B_j ] }
\ge \exp \left[ -( F_0 + N^2 \langle \tilde{S} - 
(\ln J)/N^2 -S_0 \rangle_{S_0 } ) \right]
\end{equation} 
i.e. 
\begin{equation} 
F \le F_0 + N^2 \langle \tilde{S} - (\ln J)/N^2 - S_0 \rangle_{S_0 } 
\label{linv}
\end{equation} 
with the Jacobian (which must be non-vanishing to have a 
legitimate change of variables) 
\begin{equation} 
J=\left|\frac{D\mu [M_i ] }{D\mu [B_j ] }\right| 
\end{equation}
Note that this is the determinant of a $DN^2 \times DN^2 $ matrix, where $D$
denotes the number of matrix variables.
Now, one can use the theorem of the previous section with the
immediate result that the optimal free choice of $S_0 $ in the new
variables is the free partner of $\tilde{S} - (\ln J)/N^2 $. Thus,
while for any given
set of variables, the optimal $S_0 $ is known, one may now try to further  
improve the approximation (i.e. diminish the right hand side of (\ref{linv}))
by trying different sets of variables leading to different 
$\tilde{S} $ and $J$. 

In general, this is technically difficult, since $\ln J $ is not a multinomial
expression, for which the algorithm of finding the free partner is easiest to
carry out. Therefore, in this paper the treatment will be specialized to linear
transformations, 
\begin{equation} 
M_i = \sum_{j} c_{ij} B_j
\end{equation} 
Then $J=(\det (c_{ij} ) )^{N^2 } $ is independent 
of the matrix variables $B_j $.
This greatly simplifies the treatment of (\ref{linv}); the Jacobian does not 
play any role in the determination of the optimal free action $S_0 $
for given new variables $B_j $ and thus the best free approximation
$S_0 $ to the given $\tilde{S} $ is the free partner $\tilde{S}_f $ of
$\tilde{S} $. Using furthermore that $\langle \tilde{S} -\tilde{S}_f
\rangle_{\tilde{S}_f } =0 $, one arrives at the residual minimization
problem
\begin{equation} 
F \le \tilde{F}_f - \ln J
\label{linvs}
\end{equation} 
where $\tilde{F}_f$ is the free energy associated with 
$\tilde{S}_f $. It now remains to try different choices of $c_{ij} $, 
leading to 
different $\tilde{S} $, $\tilde{S}_f $, and finally
$\tilde{F}_f$, such as to minimize the right hand side of (\ref{linvs}). 

\subsection{Calculation of the Free Energy}

Before actually carrying out the minimization procedure, it is
convenient to clarify a technical point concerning the actual
evaluation of $\tilde{F}_f $ given $\tilde{S}_f$. The best way to
achieve this seems to be the following. By subtracting a constant
$F_{ref} $, independent of $c_{ij}  $, from both sides of (\ref{linvs}),
one arrives at the equivalent problem of minimizing $\tilde{F}_f -
\ln J -F_{ref} $. For $F_{ref} $, one can choose a reference free energy,
preferably of an independent matrix model with the same number of
variables as $S$ and whose potential $V_{ref} $ is a polynomial of the same
degree as $S$. This specification is not
necessary, but convenient from the point of view of keeping the
expressions one deals with in practice as regular as possible. In
many applications, $V_{ref}(B_i ) = (1/N) \mbox{Tr} \, \sum_{i} B_i^2 $  
is adequate. Then one can write
\begin{eqnarray}
\tilde{F}_f - F_{ref} &=& 
- \ln \int \prod_j  D\mu[B_j] e^{-N^2 \tilde{S}_f[B] } 
+ \ln \int \prod_j D\mu[B_j] e^{-N^2 V_{ref} [B] } \nonumber\\ 
&=& \left. - \ln \int
\prod_j D\mu[B_j] e^{-N^2 (\alpha \tilde{S}_f[B] + (1-\alpha ) V_{ref} [B] )}
\right|^{\alpha =1 }_{\alpha   =0 } \nonumber\\ 
&=& -\int_{0}^{1} d\alpha \,
\frac{\partial}{\partial \alpha}\ln \int\prod_j D\mu[B_j]
e^{-N^2 \tilde{S}^{\prime}_f [B,\alpha] } \\  
&=& N^2 \int_{0}^{1} d\alpha \,   
\langle
\tilde{S}_f [B] - V_{ref} [B]
\rangle_{\tilde{S}^{\prime}_f }
\label{linmod}
\end{eqnarray} 
where
\begin{equation}
\tilde{S}_f^{\prime} [B,\alpha] = 
\alpha \tilde{S}_f [B] + (1-\alpha ) V_{ref} [B ] 
\label{dpridef}
\end{equation} 
In practice, the matrix model described by $\tilde{S}^{\prime}_f $ is
only insignificantly harder to solve than the one described by
$\tilde{S}_f $; on the other hand, expectation values like the ones
contained in
(\ref{linmod}) are very easy to evaluate and thus constitute a convenient way
of evaluating the right hand side of the variational principle (\ref{linvs}). 
Thus, in 
applications, the best formulation of the variational problem seems to be to
demand minimization of 
\begin{equation}
\int_{0}^{1} d\alpha \, \langle \tilde{S}_f [B] - V_{ref} [B]
\rangle_{\tilde{S}^{\prime}_f } -(\ln J)/N^2
\label{modprin}
\end{equation} 
as a function of the transformation matrix $c_{ij} $, where, to
recapitulate, $\tilde{S}^{\prime}_f $ is  given by (\ref{dpridef}) and
$\tilde{S}_f $ is the free partner of $\tilde{S} $, which in turn arises from
performing linear transformations on the original matrix variables $M_i $, i.e.
$S(M_i ) = S(c_{ij} B_j ) \equiv \tilde{S} (B_j ) $. 
Furthermore, $J=(\det (c_{ij} ) )^{N^2 } $, and
$V_{ref}[B]$ is a conveniently chosen reference potential 
independent of $c_{ij} $
generating an independent matrix model. 

As a last remark, if one has succeeded in minimizing (\ref{modprin}) by varying 
the transformation $c_{ij} $, one will usually be interested in transforming
back to the original variables; e.g., given the eigenvalue distributions of the
matrices $B_j $, one may wish to extract the eigenvalue distributions of the
matrices $M_i $. Here, the additive convolution techniques for free variables
developed in
\cite{Voi} find fruitful application, as will become clear in the examples 
to be treated further below. 

\section{Examples} \label{secexa}

\subsection{Models Involving Two Classical Matrices} 
The simplest type of interacting matrix model
involves just two classical matrices; this provides a first testing ground
for the variational approach developed in the previous sections. Without
any deeper motivation, the actions 
$$
S_1 = \frac{1}{N} \mbox{Tr} \,  (M_1^4 +M_2^4 + gM_1^2 M_2^2  )
$$ 
and 
$$
S_2 = \frac{1}{N} \mbox{Tr} \, (M_1^2  + M_2^2 + gM_1 M_2 M_1 M_2 )
$$ 
will be considered. The variational approximation will be compared
mainly with Monte Carlo experiments, although there also already exist
exact analytical solutions to some simple models of this type, see e.g.
\cite{Che} in the case of $S_2 $.

The models considered in this section are invariant under
the transformation $M_i
\rightarrow -M_i $, and throughout the treatment it will be assumed for 
simplicity
that, at the large-$N$ saddle point, $\langle M_i \rangle =0$. Of course, 
it constitutes
no problem in principle to work without this assumption and to verify
it from the
solution; this would just unnecessarily complicate the notation. In more
complicated models, it may become an interesting issue whether such a
reflection symmetry can be broken spontaneously. 

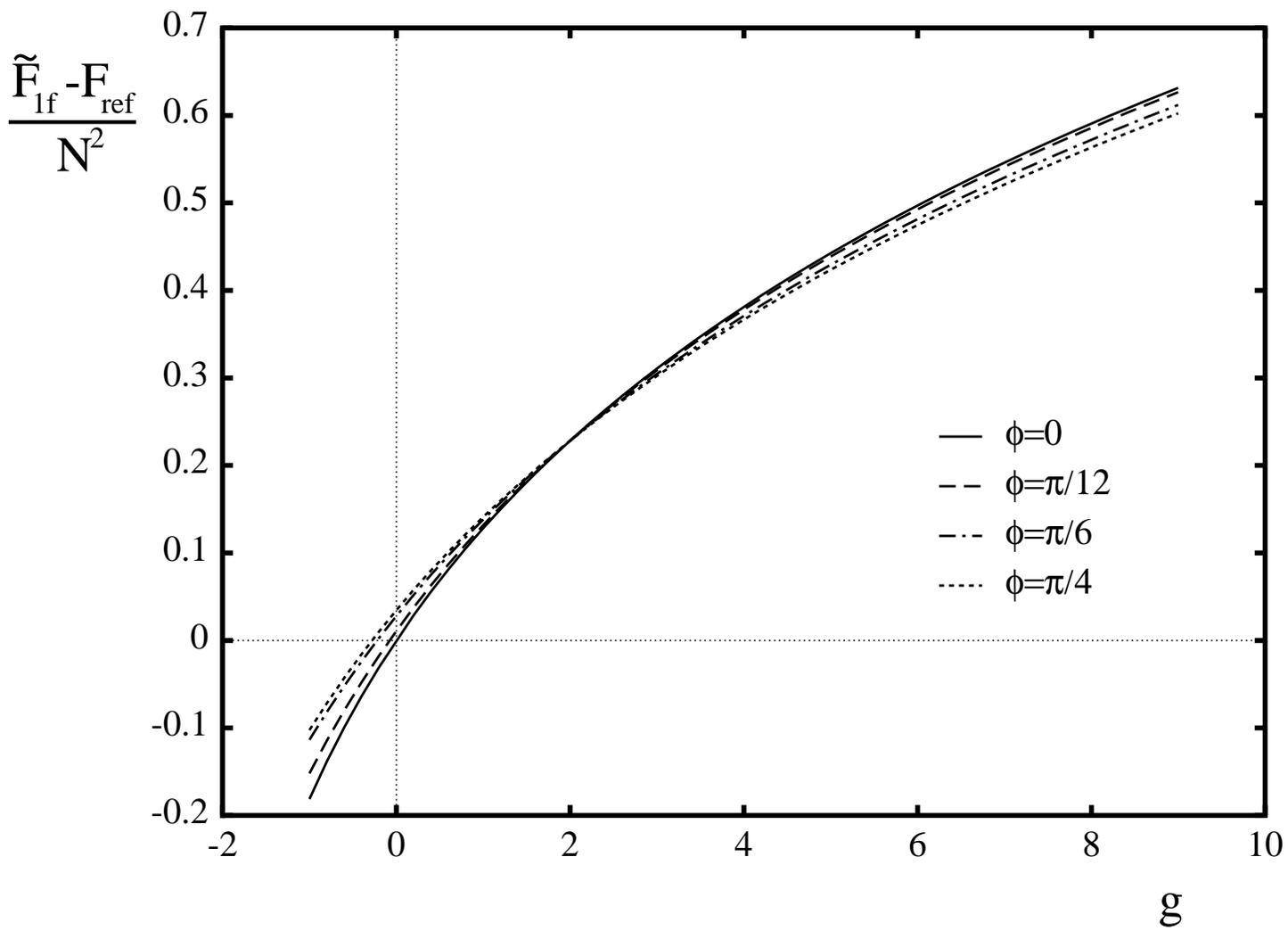
\begin{figure}
\caption{Free energies of free variational approximations to 
$S_1 $ as a function of the coupling constant $g$ for different
sets of linearly transformed variables $M_i = \sum_{j} c_{ij} B_j $,
where $c_{ij} $ is orthogonal, $c_{ij} = \sigma^{3}_{ij} \cos \phi 
+\sigma^{1}_{ij} \sin \phi $ (and $\sigma^{k} $ are the Pauli
matrices). Note how all trajectories cross at $g_{\rm cr} =2$;
$\phi =0$ provides the best approximation below $g_{\rm cr} $ and
$\phi =\pi /4 $ above $g_{\rm cr} $. Trajectories generated by
nonorthogonal $c_{ij} $ always lie higher and are not displayed.}
\label{fig0}
\end{figure}

Consider now the model described by $S_1 $. This model will be treated here in
some detail to exhibit the new techniques in practice. Later examples will
receive a more cursory treatment. Allowing for a general linear transformation 
of variables, 
\begin{equation}
M_i = \sum_{j} c_{ij} B_j
\label{glt}
\end{equation} 
and inserting into $S_1 $ to arrive at $\tilde{S}_{1} (B_j )$, one
obtains for the free partner of the latter 
\begin{equation}
\tilde{S}_{1f} = b_1 \frac{1}{N} \mbox{Tr} \, B_1^4 + 
b_2  \frac{1}{N} \mbox{Tr} \,B_2^4 + b_{12} \frac{1}{N^2  } \mbox{Tr} \,  B_1^2
\mbox{Tr} \, B_2^2
\end{equation}
with the notations
\begin{eqnarray} 
b_1 &=& c_{11}^4 + c_{21}^4 + gc_{11}^2 c_{21}^2 \nonumber \\
b_2 &=&  c_{12}^4   + c_{22}^4  + gc_{12}^2 c_{22}^2  \\
b_{12} &=&  4(c_{11}^2 c_{12}^2 +    c_{21}^2 c_{22}^2 )  +   g(c_{11}^2
c_{22}^2  + c_{12}^2 c_{21}^2 + 2 c_{11} c_{12} c_{21} c_{22} ) \; .\nonumber 
\end{eqnarray}
Here the reflection symmetry
$\langle B_j \rangle =0$ has been assumed to carry over from $\langle M_i
\rangle =0$. Choosing furthermore the reference action 
\begin{equation} 
V_{ref} = \frac{1}{N} \mbox{Tr} \, (B_1^4 + B_2^4 )
\end{equation} 
one has (cf. Eq. (\ref{dpridef}))
\begin{equation} 
\tilde{S}^{\prime}_{1f} = 
(\alpha b_1 +1-\alpha ) \frac{1}{N} \mbox{Tr} \, B_1^4 +
(\alpha b_2 +1-\alpha ) \frac{1}{N} \mbox{Tr} \, B_2^4 +
\alpha b_{12} \frac{1}{N^2 } \mbox{Tr} \, B_1^2 \, \mbox{Tr} \, B_2^2
\end{equation} 
Treating the last term in analogy to (\ref{freex}) in conjunction with 
(\ref{freeea}) and (\ref{freeeb}), one has to solve one-matrix problems for
$B_1 $ and $B_2 $ with the moments $\langle B_i^2 \rangle $ to be determined
self-consistently. The explicit solution of the one-matrix model with
arbitrary symmetric quartic potential is listed in Appendix \ref{appa}; 
the expressions for the second moments yield equations
for the $\langle B_i^2 \rangle $ which can be solved numerically for 
given $\alpha $ and $c_{ij} $. Then $\langle
\tilde{S}_{1f} -V_{ref}\rangle $ can be evaluated, again using the 
expressions for
the moments given in Appendix \ref{appa}. Finally, evaluating the integral
over $\alpha $ leads to the free energy, cf. Eq. (\ref{modprin}), with the  
additional Jacobian piece $(\ln J)/N^2 = \ln (c_{11} c_{22} - c_{12} c_{21} )$.
Minimizing the free energy numerically as a function of the $c_{ij} $ yields
the following result, cf. also Fig. \ref{fig0}:
For sufficiently small $|g|$, the original
variables $M_1 $ and $M_2 $ are the optimal ones for a free approximation,
whereas for large $g$, it  becomes favorable to switch variables by
substituting $M_1 = B_1  + B_2 $ and $M_2 = B_1 - B_2 $.
This change in behavior is of course simply induced by the competition
between the different terms in the action. For sufficiently low $g$, the 
quartic part,
which is already independent in the original variables
$M_1 $, $M_2 $, dominates. For larger $g$, it becomes favorable to accomodate 
the interaction
term as much as possible. The competition is essentially linear 
in the sense that
one optimizes the treatment of either one or the other part of the action and 
the switch between the optimal choices of variables $M_1 $, $M_2 $ or $B_1 =
(M_1 +  M_2 )/2$, $B_2 = (M_1 - M_2  )/2$ happens suddenly at $g_{\rm cr} = 
2$; the optimal choice does not change continuously with $g$.

The authors have considered the possibility that this transition 
in the variational approximation may signal a large-$N$
phase transition also in the exact solution of the model, visible e.g.
as a discontinuity of the derivative of the exact free energy as a 
function of the coupling constant $g$.
However, the Monte Carlo results do not show
such a transition -- the expectation of $\mbox{Tr} \, M_1^2 M_2^2$ 
displays a smooth behaviour
near $g_{\rm cr} $. Thus, the transition in the variational 
calculation seems to
merely reflect a crossover between two physically different regimes 
rather than a
genuine large-$N$ phase transition. A higher order phase transition can
however not be ruled out on the basis of the data taken by the authors.

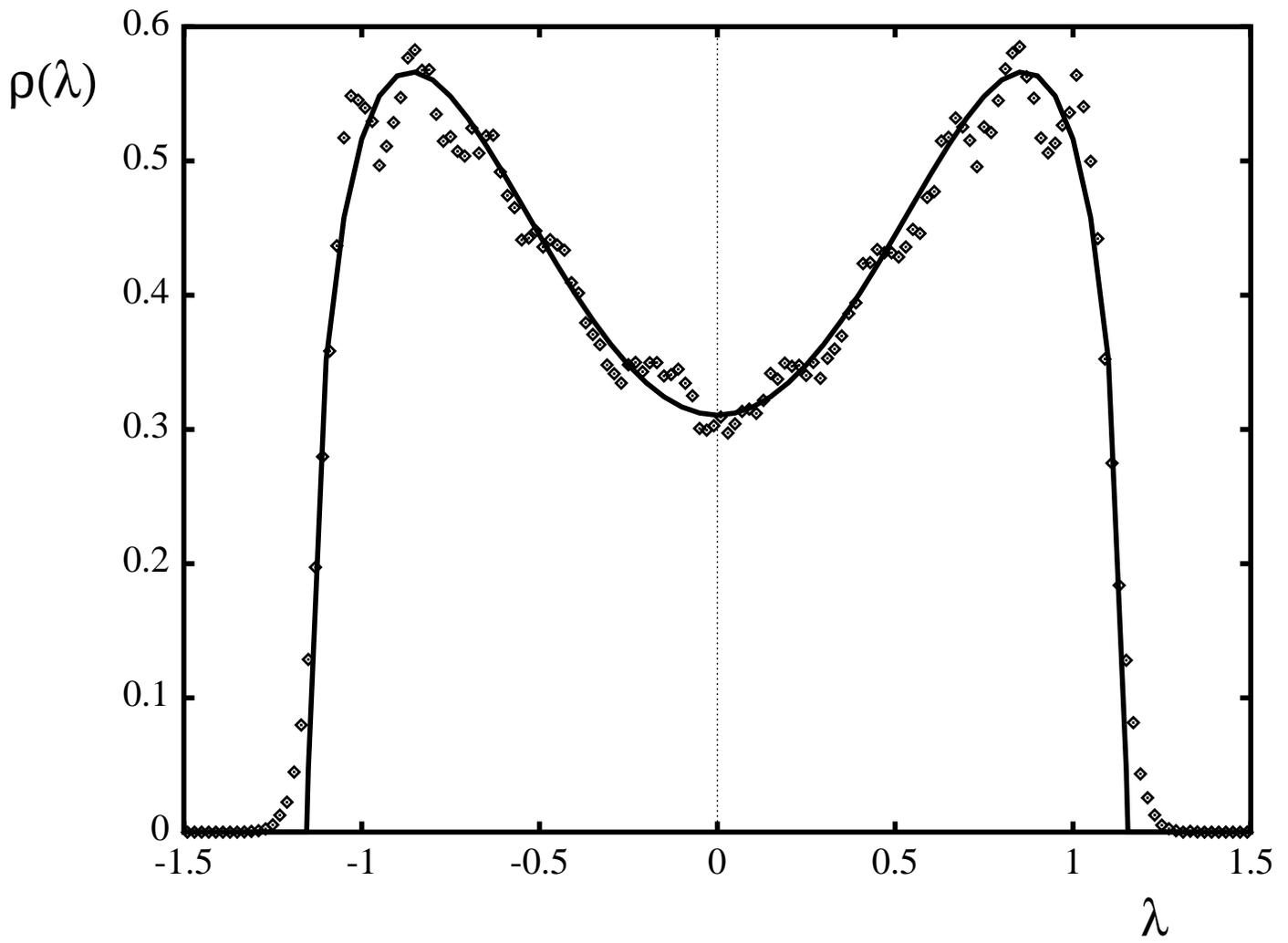
\begin{figure}
\caption{Eigenvalue distribution $\rho (\lambda )$ 
generated by $S_1 $ at $g=-1$.
Data points: Monte Carlo result for $N=10$. Solid line: Best free
variational approximation.}
\label{fig1}
\end{figure}

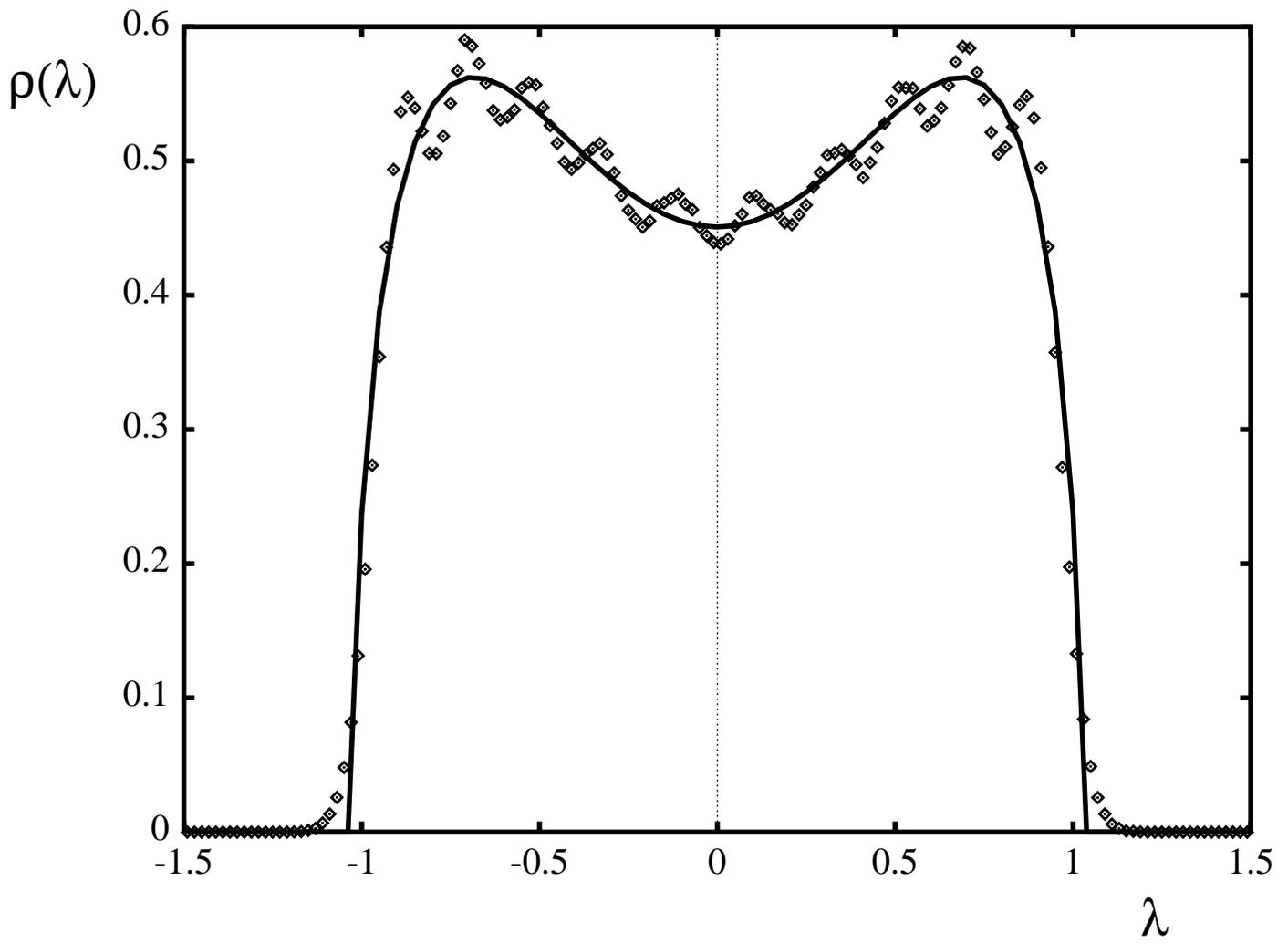
\begin{figure}
\caption{Eigenvalue distribution $\rho (\lambda )$ 
generated by $S_1 $ at $g=1$.
Data points: Monte Carlo result for $N=10$. Solid line: Best free
variational approximation.}
\label{fig2}
\end{figure}

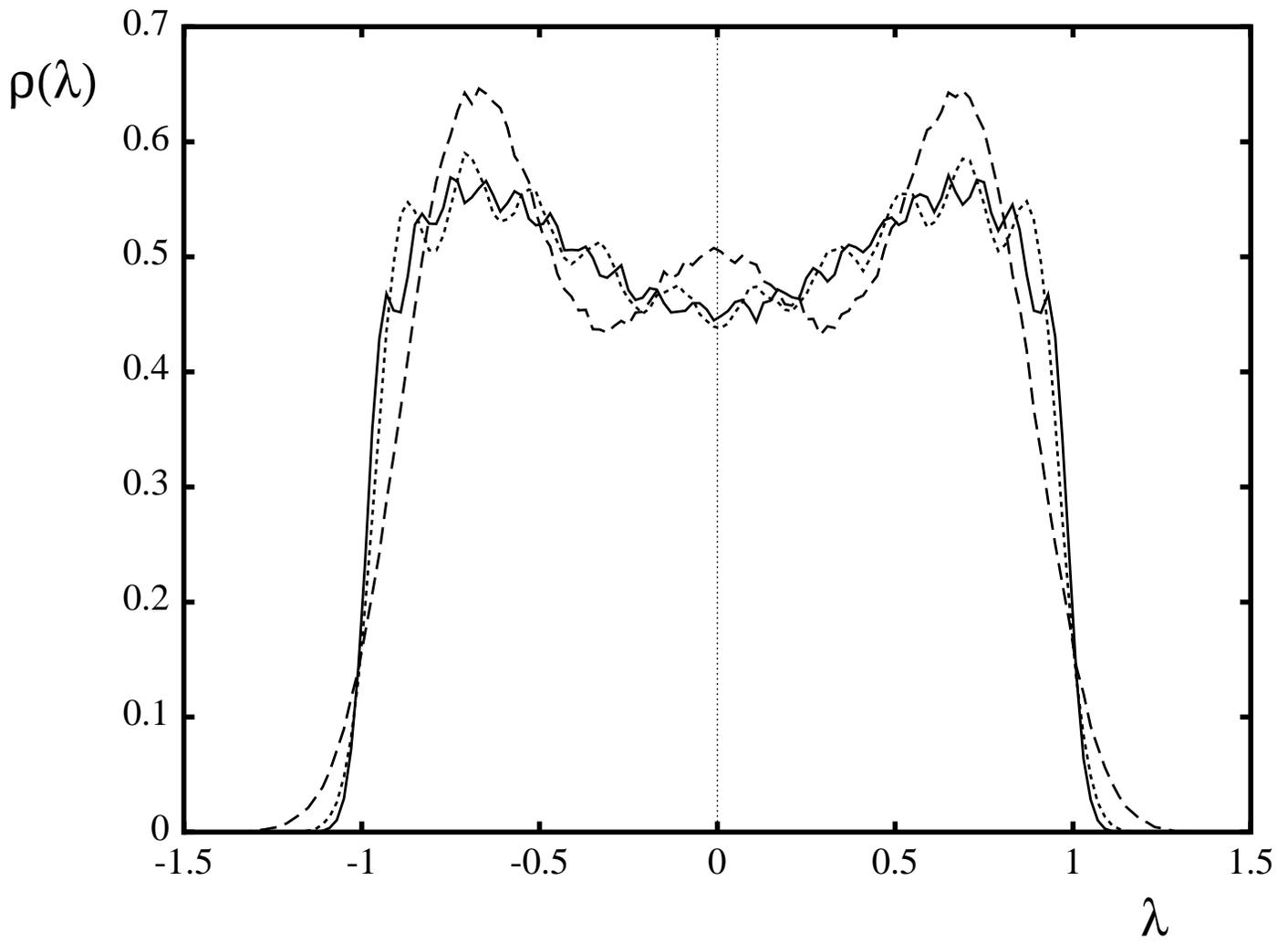
\begin{figure}
\caption{Eigenvalue distribution $\rho (\lambda )$ 
generated by $S_1 $ at $g=1$
in Monte Carlo simulations for $N=20$ (solid line), $N=10$ (dotted),
$N=3$ (dashed).}
\label{fig3}
\end{figure}

Having ascertained the optimal choice of variables, one can proceed to give the 
corresponding approximations to the eigenvalue distributions. For $g=\pm 1$,
the original variables $M_1  $, $M_2 $ are best; the free partner of $S_1 $ is 
\begin{equation} 
S_{1f} = \frac{1}{N} \mbox{Tr} \, (M_1^4 + M_2^4 ) \pm 
\frac{1}{N^2 } \mbox{Tr} \, M_1^2 \mbox{Tr} \, M_2^2 
\end{equation} 
Using the formulae of Appendix \ref{appa} to solve these models,
one obtains (assuming the eigenvalue distributions of $M_1 $ and $M_2 $ to be
identical) the consistency conditions 
\begin{equation}
\langle  M_1^2  \rangle = \langle  M_2^2  \rangle = x_M = \frac{1}{108}
\left[ \mp x_M (x_M^2 +18) + (x_M^2 +12)^{3/2} \right] 
\end{equation} 
solved by $x_M =0.334$ for $g=1$ and $x_M =0.476$ for $g=-1$ (the
relevant solution can be picked out by using positivity of $x_M $, etc.).
The corresponding eigenvalue distributions are given by (cf. Appendix 
\ref{appa})
\begin{equation}
\rho_{M} (\lambda ) = \frac{1}{\pi } (2\lambda^{2} +x_M +m^2 )
\sqrt{m^2 -\lambda^{2} } \ \ \ \ \mbox{with} \ \ \ \ 
m^2 = \frac{1}{3} (\sqrt{x_M^2 +12} -x_M )
\end{equation}
These distributions are plotted in
Figs.~\ref{fig1} and \ref{fig2}, and compared with Monte Carlo results for
$10\times 10 $ matrices. In the case of $g=1$, the dependence of the 
Monte Carlo
results on the size of the matrices is exhibited in Fig. \ref{fig3}. 
In this, as in all other cases
tested, $N=10$ already seems to embody the large-$N$ asymptotic bulk
behavior rather well, up to the characteristic oscillations in the eigenvalue
density induced  by the eigenvalue repulsion. In this respect, no deviation was
observed from the well-known finite-$N$ phenomena observed in one-matrix
models. 

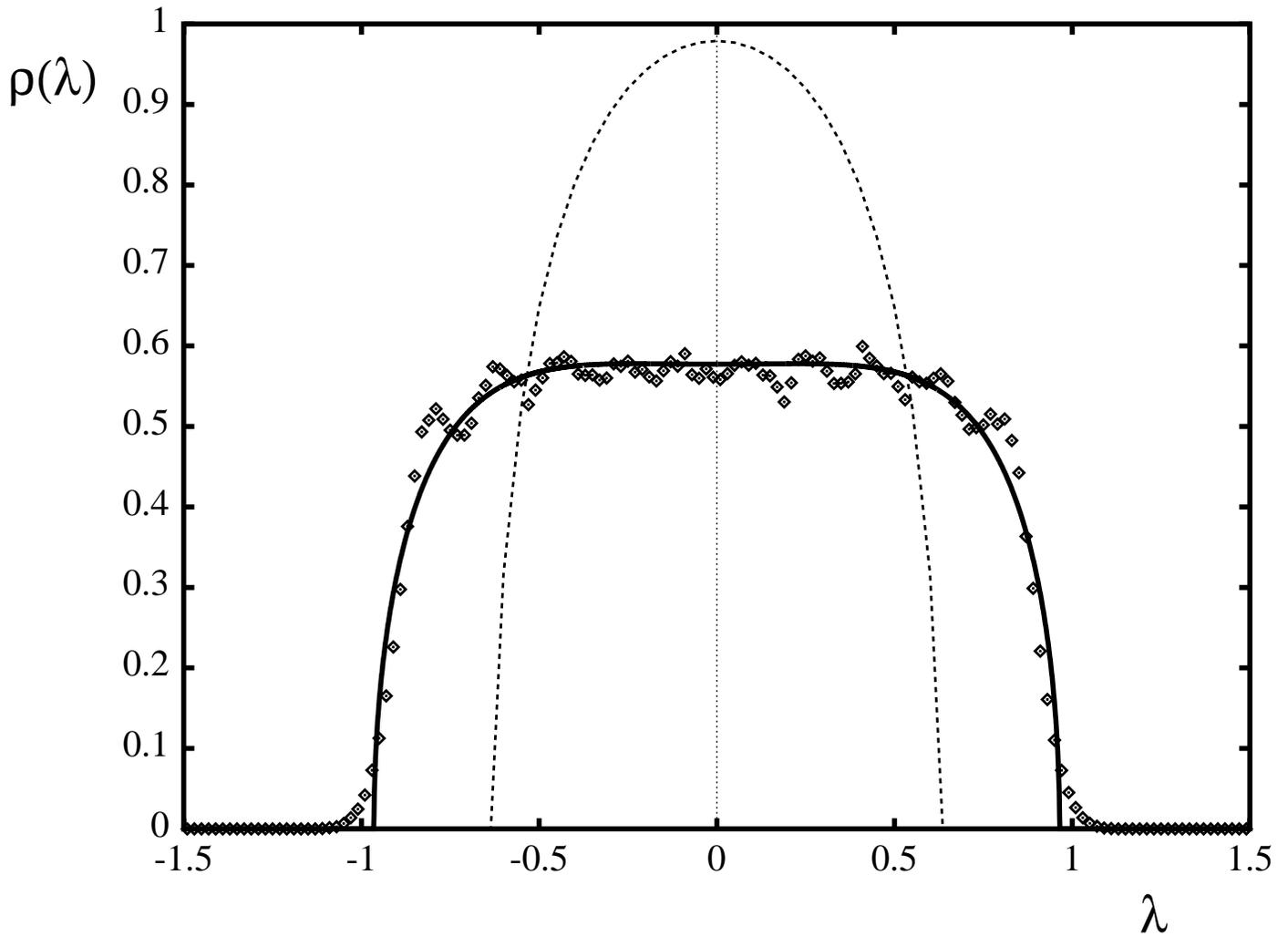
\begin{figure}
\caption{Eigenvalue distribution $\rho (\lambda )$ 
generated by $S_1 $ at $g=4$.
Data points: Monte Carlo result for $N=10$. Solid line: Best free
variational approximation, namely in the variables 
$M_1 + M_2 $, $M_1 -M_2 $. Dotted: Free approximation in the
original variables $M_1 $, $M_2 $.}
\label{fig4}
\end{figure}

For $g=4$, on the other hand, one considers the best free approximation
after the variable substitution $M_1 = B_1  + B_2 $ and
$M_2 = B_1 - B_2 $, i.e. 
\begin{equation}
\tilde{S}_{1f}   =  \frac{6}{N} \mbox{Tr}  \,    (B_1^4  + B_2^4  )  +
\frac{8}{N^2 } \mbox{Tr} \, B_1^2 \mbox{Tr} \, B_2^2 
\label{s1fjk}
\end{equation} 
Here, the consistency condition becomes
\begin{equation}
\langle  B_1^2 \rangle  =  \langle B_2^2 \rangle  = x_B = \frac{1}{486}
\left[ -4x_B (16 x_B^2 +27) + (16 x_B^2 +18)^{3/2} \right] 
\label{s1fly}
\end{equation} 
solved by $x_B =0.131$. Eqs. (\ref{s1fjk}) and (\ref{s1fly}) imply that
the $B_i $ are both governed by the quartic potential
$(1/N) \mbox{Tr} \, (6B_i^4 + 8x_B B_i^2 )$, and the corresponding
eigenvalue distribution is (cf. Appendix \ref{appa})
\begin{equation}
\rho_{B} (\lambda ) = \frac{1}{\pi } (12\lambda^{2} +8x_B +6m^2 )
\sqrt{m^2 - \lambda^{2} } \ \ \ \ \mbox{with} \ \ \ \ 
m^2 = \frac{1}{18} (\sqrt{64x_B^2 + 72} - 8x_B )
\end{equation}
In order to obtain the eigenvalue distributions of the original variables
$M_1 $ and $M_2 $ from this, one must use the free additive convolution
techniques derived in \cite{Voi}. In Appendix \ref{appb}, the convolution of two
identical distributions governed by an arbitrary symmetric quartic potential is
discussed. The resulting eigenvalue distribution $\rho_{M_1  } = \rho_{M_2  } $ 
is plotted in Fig.~\ref{fig4} along with the corresponding Monte Carlo results  
for $N=10$. Also, the comparison with the free approximation in the
original variables $M_1  $, $M_2 $ is plotted. Evidently, allowing the choice 
of free variables to vary leads to a vastly improved approximation in this case.

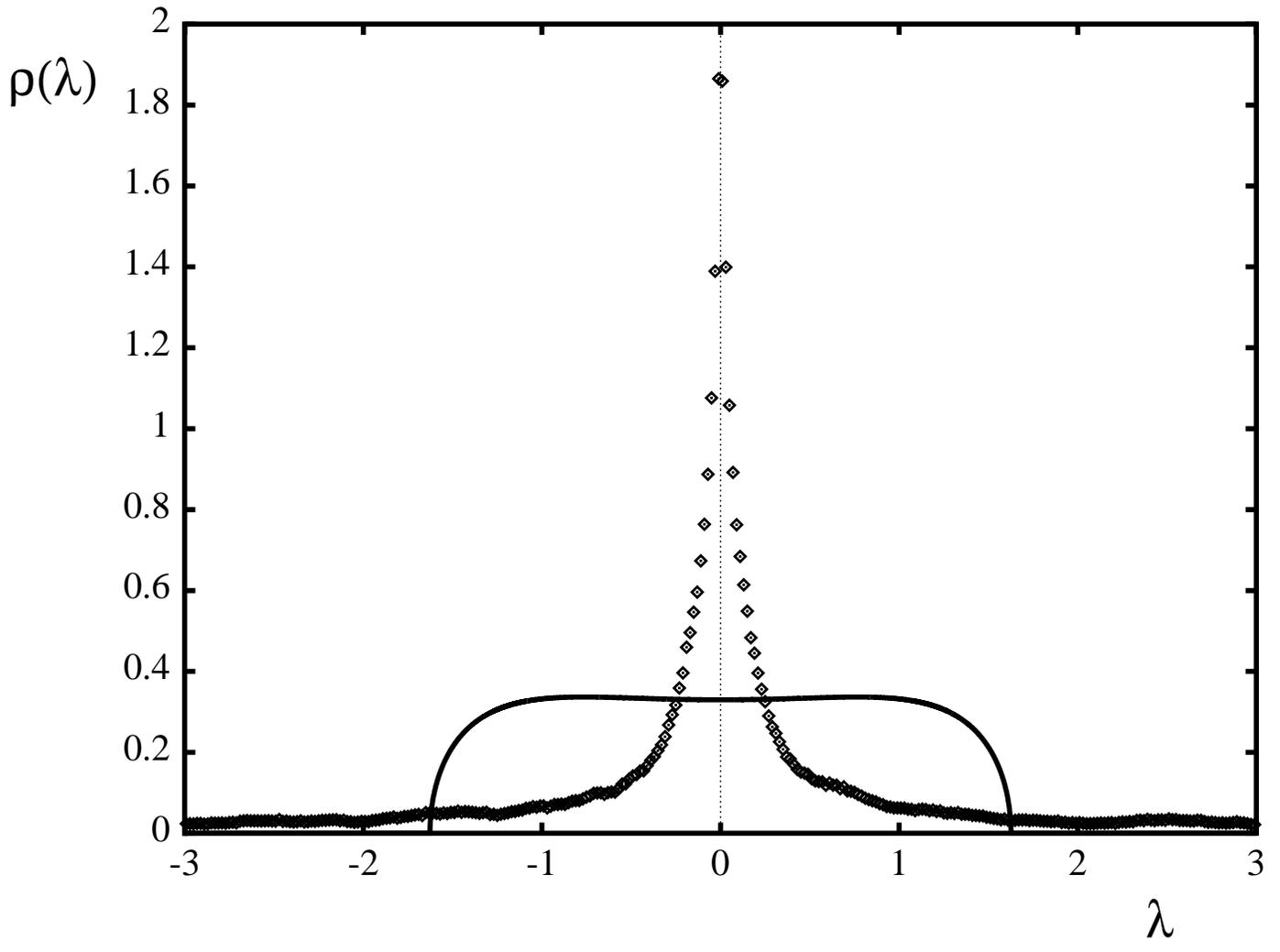
\begin{figure}
\caption{Eigenvalue distribution $\rho (\lambda )$ 
generated by $S_1^{red} $.
Data points: Monte Carlo result for $N=10$. Solid line: Best free
variational approximation.}
\label{fig5}
\end{figure}

In exactly the same vein as the case $g=4$ one can treat the extreme 
case (corresponding, up to a rescaling, to $g\rightarrow \infty $)
\begin{equation}
S_1^{red} = \frac{1}{N}
\mbox{Tr} \, M_1^2 M_2^2 \; .
\label{s1redef}
\end{equation}
Here, in the absence of any piece in the action 
which is free in the original variables and which even in the case of quite 
strong
coupling $g=4$ (see above) kept the behavior of the model reasonably regular, 
the
disagreement between the variational calculation and the exact result is rather
catastrophic, cf. Fig.~\ref{fig5}. This is not hard to understand; in the case 
of the action $S_1^{red} $, the variables $M_1 $, $M_2 $ can use
configurations such as 
\begin{equation} 
M_1 = \left( \begin{array}{cc} m_1 & 0 \\ 0 & 0 \end{array}
\right)
\ \ \ \ \ \ \ \ \ \ \ \ \ 
M_2 = \left( \begin{array}{cc} 0 &  0 \\ 0 & m_2 \end{array}
\right) 
\label{redco}
\end{equation} 
where $m_1 $, $m_2 $ are roughly $N/2 \times N/2 $ matrices, to
preserve a low value of $S_1^{red} $ while realizing very large eigenvalues
(together with a concentration of very small ones). This is indeed what is seen
in Fig.~\ref{fig5}. Dominance of configurations such as (\ref{redco})
implies a strong angular correlation between the matrices; if one rotates the
above matrices with respect to one another, e.g. such that the eigenvalues are
arbitrarily reordered, $S_1^{red} $ will in general take a very large value.
Such 
angular correlations are completely lost in the free partner $S_{1f}^{red} =
(1/N^2 ) \mbox{Tr} \, M_1^2 \mbox{Tr} \, M_2^2 $ and evidently even the best
choice of linearly transformed variables $M_1 = B_1 + B_2  $ and $M_2 = B_1
-B_2 $, though incorporating some angular correlations between $M_1 $ and
$M_2 $, cannot do much to alleviate this problem. It is however
possible that a nonlinear change of variables exists for which the
difficulty will be satisfactorily removed. The authors did not explore
this direction. The dismal failure of the
variational approximation in this case thus does not come as a complete
surprise. Rather, it is gratifying to see how well the variational 
approximation still
works in the previous, not quite as pathological, example described by $S_1 $ 
with
the strong coupling of $g=4 $. One striking feature of the Monte Carlo 
eigenvalue
distribution in the case of $S_{1}^{red} $ is the absence of the characteristic
finite-$N$ oscillations observed in all other cases. This can be understood 
as follows:
Usually, the potential confines the eigenvalues to a compact domain and, due to
the eigenvalue repulsion, they tend to be equidistant. This (fluctuating) 
lattice has a favored equilibrium position, leading to the oscillations in   
the eigenvalue density. However, when the action allows the eigenvalues to  
spread over the entire real axis, the correlations induced by the eigenvalue
repulsion become insignificant and the oscillations in the eigenvalue density
disappear\footnote{The authors are indebted to U.-J. Wiese for this remark.}. 

Apart from plotting the eigenvalue distributions, which essentially contain the
information about all the moments of the individual matrices, one can test 
mixed correlators for freeness properties. In particular, if the matrices 
$M_1 $ and
$M_2 $ are free with respect to each other, then there are e.g. the following
relations between correlators, following from the axioms of freeness: 
\begin{eqnarray} 
C_1 \equiv \langle M_1^2 M_2^2 \rangle &=& 
\langle M_1^2 \rangle \langle M_2^2 \rangle \equiv C_{1f} \nonumber \\ C_2 
\equiv   \langle M_1^4 M_2^4  \rangle  &=&  \langle M_1^4 \rangle
\langle M_2^4 \rangle \equiv C_{2f} 
\label{mixmf} \\ C_3 \equiv \langle M_1^2 M_2^2 M_1^2 M_2^2 \rangle &=& 
\langle M_1^4   \rangle  \langle M_2^2  \rangle^{2}  +  \langle  M_1^2
\rangle^{2} \langle M_2^4 \rangle -  \langle M_1^2 \rangle^{2} \langle M_2^2
\rangle^{2} \equiv C_{3f} \nonumber 
\end{eqnarray} 
In the Monte Carlo calculations, both sides of these equations were
sampled in order to give a measure for the deviation from freeness 
(i.e. strength 
of angular correlations) contained in the exact models. The results are 
tabulated
in Table~1 for different values of
$g$ in $S_1 $. Note that for $S_1^{red} $, the moments of the exact distribution
diverge due to the evidently too slow fall-off of the eigenvalue distribution
for large eigenvalues, making such a comparison impossible. 

\begin{center}
\begin{tabular}{|c||c|c||c|c||c|c||c|c|}
\hline \rule[-1.5ex]{0ex}{4.5ex} $g$  &   $\langle M_i^2 \rangle  $  &
$\langle M_i^4 \rangle $ & $C_1 $ &  $C_{1f} $ &  $C_2 $ & $C_{2f} $ &
$C_3 $ & $C_{3f} $ \\ \hline \hline -1 & .483 & .377 & .254 & .233 & .171 & .142
& .150 & .121 \\ \hline 1 & .336 & .197  & .106 & .113 & .0339 & .0387  & .0272
& .0317 \\  \hline 4 & .267 & .133 & .0585 & .0713 & .0116 & .0177 & .0085 &
.0139 \\ \hline 
\end{tabular} \vspace{0.3cm}

Table 1 : Mixed correlators, Eqs. (\ref{mixmf}), for the model $S_1$ compared to
the predictions for free $M_i$ 
\end{center}

One observes how, for growing $g$, the interaction term introduces stronger
angular correlations, evident in the stronger deviations from free predictions.
Now,
for sufficiently large $g$, e.g. $g=4$, the variational principle asserts 
that a better
approximation is obtained with a new set of mutually free variables $B_i $. 
Using
this new set of free variables, of course also the predictions for mixed
correlators, derived in (\ref{mixmf}) for free $M_i  $, change. This shall be
illustrated here for the correlator $\langle M_1^2 M_2^2
\rangle $. Inserting $M_1 = B_1 + B_2 $ and $M_2 = B_1  - B_2 $, 
using the axioms
of freeness on $B_1 $ and $B_2 $ together with $\langle B_i
\rangle =0$, and then substituting back $B_1 = (M_1 +  M_2 )/2$ and
$B_2 = (M_1 - M_2 )/2$, one obtains 
\begin{eqnarray}
\langle  M_1^2 M_2^2 \rangle   &=&  \langle B_1^4  +  B_2^4 \rangle  =
\frac{1}{8} \langle M_1^4 +  M_2^4 +4 M_1^2 M_2^2  + 2 M_1 M_2 M_1 M_2
\rangle \\ \langle M_1 M_2  M_1 M_2 \rangle  &=& \langle B_1^4 + B_2^4
\rangle -4 \langle B_1^2 \rangle \langle B_2^2 \rangle = \langle M_1^2 M_2^2 
\rangle -  \frac{1}{4}  (\langle M_1^2  +  M_2^2 \rangle^{2} -4
\langle M_1 M_2  \rangle^{2} ) \\ \langle M_1  M_2 \rangle &=& \langle B_1^2
\rangle - \langle B_2^2 \rangle =0 
\end{eqnarray} 
where in the last line, identical distributions for the $B_i $ were
assumed. Putting these relations together yields $\langle M_1^2 M_2^2
\rangle $ in terms of the individual moments, 
\begin{equation}
\langle   M_1^2  M_2^2 \rangle =  \frac{1}{2}   \langle M_1^4  + M_2^4
\rangle - \frac{1}{4} \langle M_1^2 + M_2^2 \rangle^{2} 
\label{mixbf}
\end{equation} 
In the case $g=4$, the right hand side, using the individual moments
from the Monte Carlo experiment, takes the  value .0617, which indeed is 
closer to
the exact value for $\langle  M_1^2 M_2^2 \rangle \equiv C_1 $ 
than the prediction
using free variables $M_i $ quoted in Table 1. 

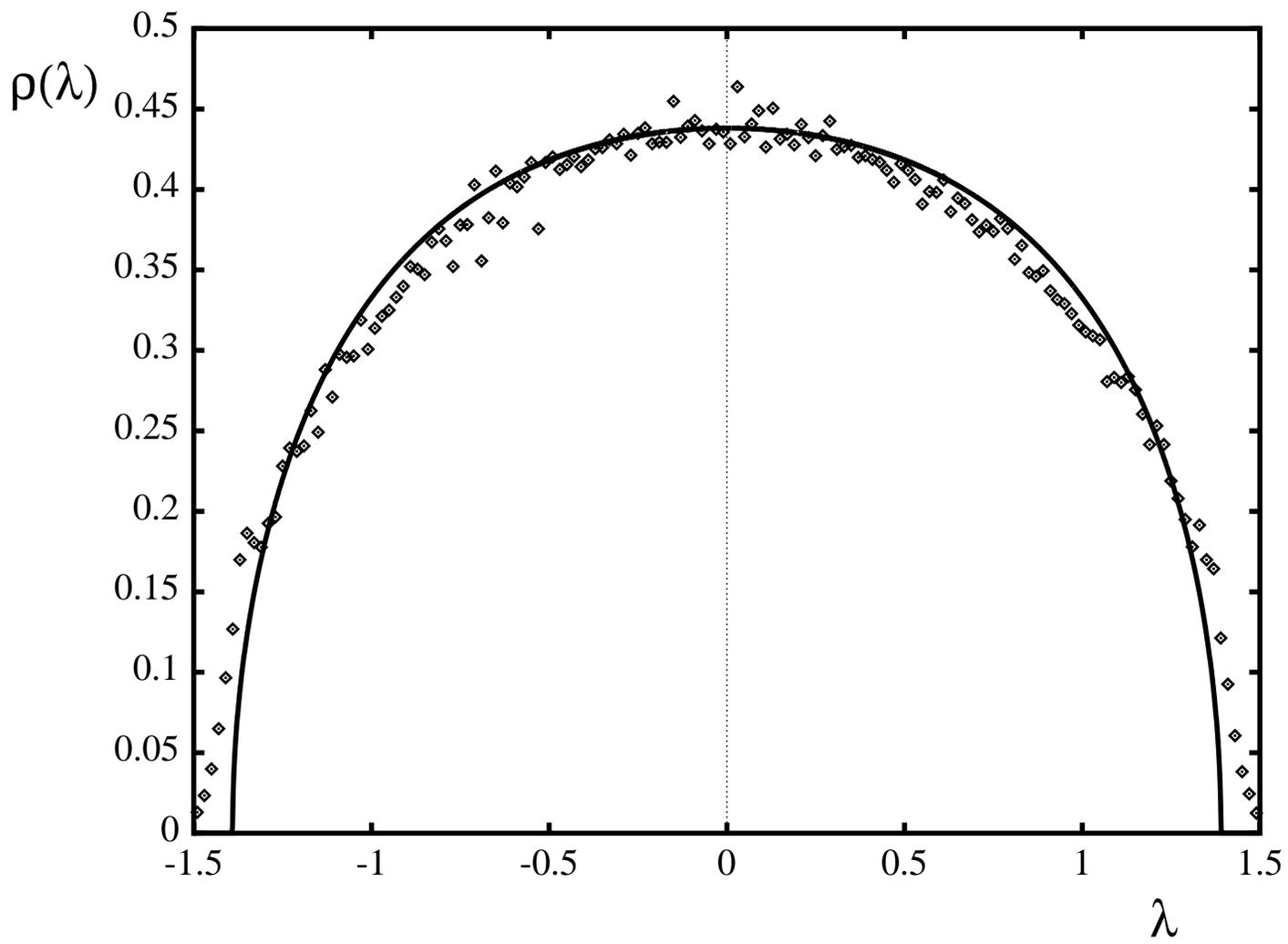
\begin{figure}
\caption{Eigenvalue distribution $\rho (\lambda )$ 
generated by $S_2 $ at $g=2/5$.
Data points: Monte Carlo result for $N=40$. Solid line: Best free
variational approximation.}
\label{fig6}
\end{figure}

Consider now the model described by 
$$
S_2 = \frac{1}{N} \mbox{Tr} \, (M_1^2 + M_2^2 + 
gM_1   M_2 M_1 M_2 ) \; .
$$
After performing a general linear transformation as in
(\ref{glt}), one arrives at the free partner 
\begin{eqnarray}
\tilde{S}_{2f} = (c_{11}^{2} + c_{21}^{2} ) \frac{1}{N} \mbox{Tr} \, B_1^2 + 
(c_{12}^{2} + c_{22}^{2} ) 
\frac{1}{N}  \mbox{Tr}  \, B_2^2 + gc_{11}^{2} c_{21}^{2} 
\frac{1}{N} \mbox{Tr}  \, B_1^4 + \nonumber \\  
+ gc_{12}^{2} c_{22}^{2} \frac{1}{N} 
\mbox{Tr}
\, B_2^4   + 4gc_{11} c_{12}  c_{21} c_{22} 
\frac{1}{N^2 } \mbox{Tr} \, B_1^2 \mbox{Tr}
\, B_2^2  \; .
\label{ts2f}
\end{eqnarray}
Evidently, from the point of view of the variational approximation,
this model is slightly simpler than the one considered above. Since the
noninteracting part of $S_2 $ is invariant under linear transformations of
the variables and subsequent construction of the free partner, up to trivial
variable rescalings (note also that, as before, $\langle B_i \rangle =0 $ was
used in deriving (\ref{ts2f})), the choice of variables will always be such 
as to best
accomodate the interacting term, for any coupling $g$. Carrying out the 
calculation of the free energy in complete analogy to the calculation 
for $S_1 $,
this turns out to be the choice $M_1 = B_1 + B_2 $, $M_2 = B_1 - B_2 $,
corresponding to the free partner 
\begin{equation}
\tilde{S}_{2f} = 
\frac{1}{N} \mbox{Tr} \, ( gB_1^4 + gB_2^4 + 2B_1^2 + 2B_2^2 ) -
\frac{4g}{N^2 } \mbox{Tr} \, B_1^2 \mbox{Tr} \, B_2^2 
\label{ts2fopt}
\end{equation} 
The corresponding consistency condition is
\begin{equation}
\langle B_1^2 \rangle = \langle B_2^2  \rangle = x_B = \frac{1}{108g^2 }
\left[ -(2-4gx_B )^3 -18g(2-4gx_B )+((2-4gx_B )^2 +12g)^{3/2} \right] 
\label{ts2foptc}
\end{equation}
From the resulting eigenvalue distributions of the matrices $B_i $
one again obtains the eigenvalue distributions of the original variables
$M_i $ by additive convolution. The result for $g=2/5$, leading to
$x_B = .2522 $ in (\ref{ts2foptc}), is displayed in  
Fig.~\ref{fig6}, compared with the Monte Carlo result for $N=40$. 
Also, various correlators are tabulated in Table 2.

\begin{center}
\begin{tabular}{|c||c|c||c|c||c|c||c|c|}
\hline \rule[-1.5ex]{0ex}{4.5ex} $g$ &
$\langle M_i^2 \rangle $ & $\langle M_i^4 \rangle $ &
$C_1 $ & $C_{1f} $ & $C_2 $ & $C_{2f} $ & $C_3 $ & $C_{3f} $ \\
\hline \hline
2/5 & .53 & .56 & .30 & .28 & .37 & .32 & .28 & .24 \\ \hline
\end{tabular} \vspace{0.3cm}

Table 2 : Mixed correlators, Eqs. (\ref{mixmf}), for the model $S_2$,
compared to the predictions for free $M_i$

\end{center}

The exact eigenvalue distribution turns out to be surprisingly similar 
to the semicircle of radius $\sqrt{2} $ obtained at $g=0$. This
is well reproduced by the variational calculation
despite a quite strong dependence on $g$ of the individual coefficients
in $\tilde{S}_{2f} $. The effects of the $g$-dependence of the 
quadratic and quartic coefficients nearly cancel in the complete 
potential, leading to a quite stable eigenvalue distribution.

Turning to the mixed correlators in Table 2, one does not observe a
strong deviation from free behavior of the $M_i $. In this case,
there is almost no room for improvement by assuming free $B_i $
instead of $M_i $, leading to equation (\ref{mixbf}) for
$\langle M_1^2 M_2^2 \rangle $. Indeed, one here obtains for the
right hand side of (\ref{mixbf}) the value $.28$, the same as
$C_{1f} $ in Table 2.

The reader may wonder why only the relatively small value $g=2/5$
was displayed in the comparison above. In fact, the model described
by $S_2 $ becomes unstable for too large values of $|g|$. Note that
the product of two hermitian matrices $M_1 M_2 $ is in general not
hermitian; therefore, the interaction term, proportional to
$(M_1 M_2 )^2 $, is not bounded from below even
for positive $g$. Strictly speaking, the model is unstable for all $g$,
but for sufficiently small $g$, there is effectively a barrier posed by
the independent part of the action preventing the system from spilling over
to the region of large eigenvalues where the interaction will allow
eigenvalues to grow without bound. A glimpse of this instability
(for negative $g$, namely setting in at $g=-4/9$) was already given in
\cite{Che}\footnote{Presumably, the model described by $S_1 $
will display a similar instability when the coupling $g$ becomes too
negative. This was not investigated further by the authors.}.
Here, instead the instability at positive $g$ was investigated
in slightly more detail. According to Monte Carlo simulations at
$N=10$, $20$, and $40$, the critical coupling lies in the interval
$g \in [0.4,0.45]$. On the other hand, the variational approximation
in the best basis, described by the action (\ref{ts2fopt}), is stable 
up to $g=2$. Beyond this point, there is no solution to the consistency
condition (\ref{ts2foptc}). The authors also checked that there is no
two-cut solution to (\ref{ts2fopt}) above $g=2$ (note that (\ref{ts2foptc})
is valid only for one-cut solutions). Thus, the variational approximation
does qualitatively capture the phase structure of the model described
by $S_2 $, albeit with a badly overestimated critical coupling. In
this respect, the choice of variables $B_1 = (M_1 + M_2 )/2$ 
and $B_2 = (M_1 - M_2 )/2$ represents a drastic improvement
over the original variables $M_i $. In terms of the latter, the free
counterpart of $S_2 $ is simply 
$S_{2f} = (1/N) \mbox{Tr} \, (M_1^2 + M_2^2 )$,
which entirely misses the unstable phase of the model (in fact, all the
dependence on the coupling $g$).

\subsection{Classical Constant Matrices in Many Dimensions}
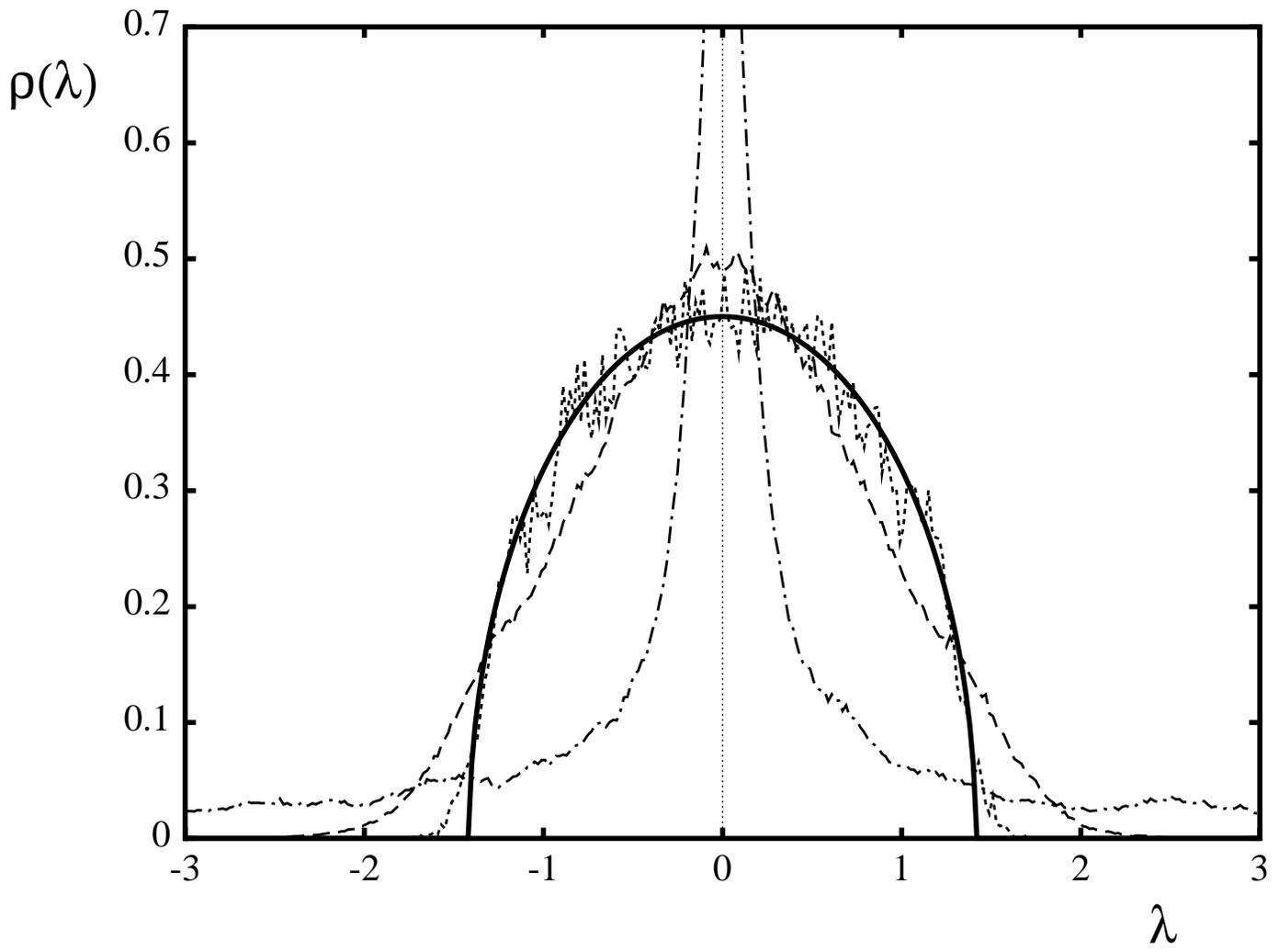
\begin{figure}
\caption{Eigenvalue distribution $\rho (\lambda )$ generated by $S_{D2} $.
Monte Carlo results for $N=10$ are given in the cases
$D=2$ (dash-dotted), $D=3$ (dashed) and $D=10$ (dotted).
Solid line: Free variational approximation in the original variables
$M_i $.}
\label{fig7}
\end{figure}

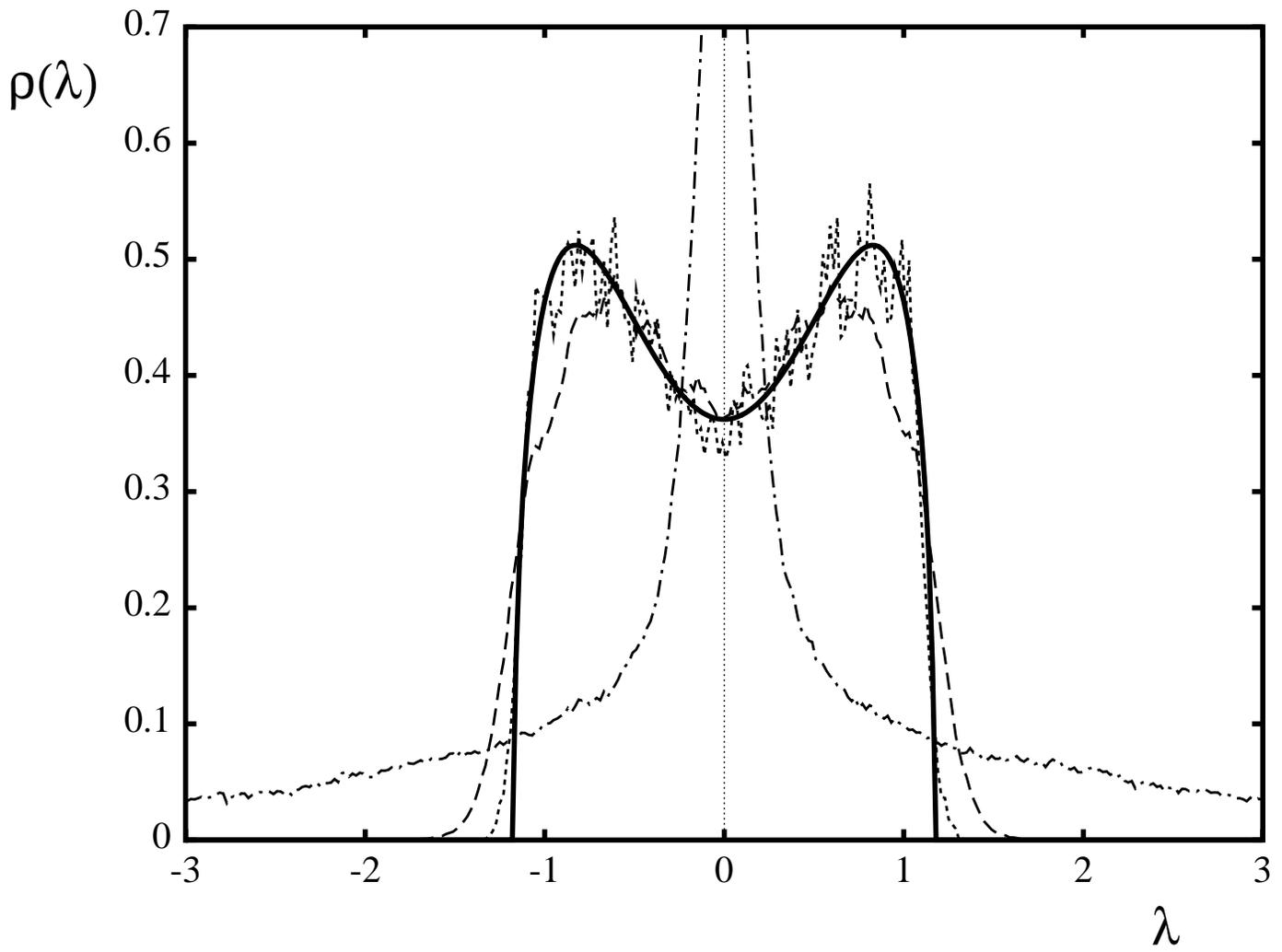
\begin{figure}
\caption{Eigenvalue distribution $\rho (\lambda )$ generated by $S_{D4} $.
Monte Carlo results for $N=10$ are given in the cases
$D=2$ (dash-dotted), $D=3$ (dashed) and $D=10$ (dotted).
Solid line: Free variational approximation in the original variables
$M_i $.}
\label{fig8}
\end{figure}

Consider a model of $D$ classical matrices with an action of the type
\begin{equation}
S_D = \frac{1}{D-1} \sum_{i\neq j}^{D} S (M_i , M_j )
\label{ldtyp}
\end{equation}
Note the scaling of this action with $D$, which is necessary to retain
a meaningful balance between it and the eigenvalue repulsion originating
in the Haar measure of the $M_i $. Before proceeding, a comment is in order
regarding the special form of (\ref{ldtyp}) and the interpretation
of $D$ as the number of space-time dimensions. In (\ref{ldtyp}), each matrix
degree of freedom interacts with all others in the same way. If one wishes
to attach physical meaning to the matrices as degrees of freedom living
in space-time, this occurs most naturally when all the degrees of freedom
are attached to the same space-time point. Such a model may become
relevant if one has managed to decouple the different space-time points
of a $D$-dimensional large-$N$ field theory,
e.g. as the result of an Eguchi-Kawai reduction (for a review, see
Ref. \cite{Egu}). Now, in view of the scaling
of the individual terms in (\ref{ldtyp}), one might hope that the detailed
angular correlations between pairs of matrices become unimportant at large
$D$ and a free approximation becomes exact. This argument can be made
rigorous e.g. in the Kazakov-Migdal model, where the leading
large-$D$ behavior is described by a free matrix model \cite{Mak1}.
The proof however depends on the specific link variable structure
of the Kazakov-Migdal model. In order to test this idea more generally,
the models described by
\begin{equation}
S_{D2} = \frac{1}{D-1} \sum_{i\neq j}^{D} 
\frac{1}{N} \mbox{Tr} \, M_i^2 M_j^2 
\ \ \ \ \ \ \ \mbox{and} \ \ \ \ \ \ \ 
S_{D4} = \frac{1}{D-1} \sum_{i\neq j}^{D} 
\frac{1}{N} \mbox{Tr} \, M_i^4 M_j^4
\label{ladmo}
\end{equation}
were investigated in the present framework for different $D$. The
reader is reminded that in the case of $S_{D2} $ for $D=2$, the
free variational approximation failed miserably (cf. previous section).
In analyzing the models (\ref{ladmo}) in the free variational
approximation, the authors did not treat different choices of free
variables; for simplicity, only the best free approximation in the
original variables $M_i $ was considered. Then the best free
approximation is easily derived: The free counterparts of (\ref{ladmo})
are
\begin{equation}
S_{D2f} = \frac{1}{D-1} \sum_{i\neq j}^{D} 
\frac{1}{N^2 } \mbox{Tr} \, M_i^2 \mbox{Tr} \, M_j^2
\ \ \ \ \ \ \ \mbox{and} \ \ \ \ \ \ \
S_{D4f} = \frac{1}{D-1} \sum_{i\neq j}^{D} 
\frac{1}{N^2 } \mbox{Tr} \, M_i^4 \mbox{Tr} \, M_j^4
\end{equation}
and thus each variable $M_i $ is described by the action
\begin{equation}
\bar{S}_{D2} (M_i ) = 2 x_{D2} \frac{1}{N} \mbox{Tr} \, M_i^2
\ \ \ \ \ \ \ \mbox{or} \ \ \ \ \ \ \
\bar{S}_{D4} (M_i ) = 2 x_{D4} \frac{1}{N} \mbox{Tr} \, M_i^4
\end{equation}
independent of $D$, with $x_{D2} = \langle M_i^2 \rangle $ and
$x_{D4} = \langle M_i^4 \rangle $ to be determined self-consistently.
One obtains $x_{D2} = 1/2 $ and $x_{D4} = 1/\sqrt{8} $ and therefore the
eigenvalue distributions
\begin{equation}
\rho_{2} (\lambda ) = \frac{1}{\pi } \sqrt{2-\lambda^{2} }
\ \ \ \ \ \ \ \mbox{and} \ \ \ \ \ \ \
\rho_{4} (\lambda ) = \frac{1}{\pi } \left( \sqrt{2} \lambda^{2} +
\sqrt{\frac{\sqrt{8} }{3} } \right) \sqrt{\sqrt{\frac{2\sqrt{8} }{3} }
-\lambda^{2} }
\end{equation}
Comparing this with the results of Monte Carlo experiments, 
cf. Figs. \ref{fig7} and \ref{fig8}, the correspondence is quite
satisfying already at $D=10$ (the simulations were carried out
for $N=10$, some checks for $N=20$ did not reveal any significant
differences).
Also, comparing mixed correlators of two matrices (without loss of
generality, $M_1 $ and $M_2 $) with free predictions 
(cf. equations (\ref{mixmf})) in the
Monte Carlo experiments, one obtains good
agreement for $D=10$, cf. Table 3.

\begin{center}
\begin{tabular}{|c||c|c||c|c||c|c||c|c|}
\hline \rule[-1.5ex]{0ex}{4.5ex} &
$\langle M_i^2 \rangle $ & $\langle M_i^4 \rangle $ &
$C_1 $ & $C_{1f} $ & $C_2 $ & $C_{2f} $ & $C_3 $ & $C_{3f} $ \\
\hline \hline
$ S_{D2}, D=3 $ &
.62 & 1.00 & .25 & .38 & .33 & 1.00 & .18 & .62 \\ \hline
$ S_{D2}, D=10 $ &
.51 & .53 & .25 & .26 & .25 & .28 & .18 & .21 \\ \hline \hline
$ S_{D4}, D=3 $ &
.49 & .46 & .20 & .25 & .12 & .21 & .09 & .16 \\ \hline
$ S_{D4}, D=10 $ &
.46 & .37 & .21 & .22 & .13 & .13 & .11 & .11 \\ \hline
\end{tabular} \vspace{0.3cm}

Table 3 : Mixed correlators, Eq. (\ref{mixmf}), for the
models (\ref{ladmo}), and predictions for free $M_i $  
\end{center}

As mentioned above, this agreement does not come unexpected.
However, while in the cases considered here the Monte Carlo solution
seemed to converge towards the free approximation at large
$D$, it seems doubtful that this should be true in general. 
On the one hand, a solution with angular correlations, say 
with angles restricted to a certain sector as compared with some fixed
reference matrix, costs a weight
$\exp (-\alpha N^2 D )$ in the partition sum (there are $O(N^2 D )$ angles, 
and the
coefficient $\alpha $ depends on the details of the angular distributions). 
On the 
other hand, the interaction itself is also scaled to be of order $O(N^2 D )$.
Therefore,
there should be a genuine competition between the two terms, the former
favoring freeness and the latter preventing it. 

\subsection{Quantum Mechanics of One Matrix} 
For a wide class of potentials, the
quantum mechanical ground state of a single large-$N$ matrix can be solved for
exactly \cite{Bre1}. This is achieved by working directly with continuous
(Euclidean) time and mapping the problem to a noninteracting fermion 
problem. If one discretizes time as in the original definition of the path
integral, the problem can be interpreted in terms of many interacting classical
matrices. In this form, the quantum mechanics of one matrix is hard to solve
exactly; on the other hand, it becomes amenable to the variational method dealt
with here. Interest in such a discretized form has arisen in recent 
applications to quantum gravity, \cite{GroK}.
For definiteness, the model described by 
\begin{equation} 
S_{Q1} = \frac{1}{N} \mbox{Tr} \, \left( \sum_{i=0}^{L-1} (M_{i+1}
-M_i )^2 + gM_i^4 \right) 
\label{sq1def}
\end{equation} 
will be considered here, where $L$ is the number of time slices, and
$M_0 $ is identified with $M_L $, i.e. periodic boundary conditions
are posited. Generically, there is a competition between two different
interaction terms: The nearest-neighbor coupling and the potential term,
which acts instantaneously. In order to best accomodate the former in
a free variational approximation, one will want to change matrix
variables to a Fourier basis, which decouples the kinetic part of the
action. On the other hand, the potential term will prefer the original
variables. Thus, in the present treatment, not a general linear change 
of variables will be considered, but only the aforementioned two discrete
choices; the variational minimization principle will decide which choice
is preferred. The Fourier basis is defined as
\begin{equation}
M_i = \sum_{j} U_{ij} B_j
\label{foutra}
\end{equation}
with ($L$ is for simplicity taken to be even)
\begin{equation}
U = \sqrt{\frac{2}{L} } \left( \begin{array}{cccccccc}
0 & \cdots & 0 & 1/\sqrt{2} & 1 & \cdots & 1 & 1/\sqrt{2} \\
\sin k_{L/2 -1} & \cdots & \sin k_{1} & 1/\sqrt{2} & \cos k_{1} & \cdots &
\cos k_{L/2 -1} & -1/\sqrt{2} \\
\sin 2k_{L/2 -1} & \cdots & \sin 2k_{1} & 1/\sqrt{2} & \cos 2k_{1} & \cdots &
\cos 2k_{L/2 -1} & 1/\sqrt{2} \\
\sin 3k_{L/2 -1} & \cdots & \sin 3k_{1} & 1/\sqrt{2} & \cos 3k_{1} & \cdots &
\cos 3k_{L/2 -1} & -1/\sqrt{2} \\
\vdots & \ddots & \vdots & \vdots & \vdots & \ddots & \vdots & \vdots \\
\sin (L-1)k_{L/2 -1}& \cdots & \sin (L-1)k_{1} & 1/\sqrt{2} &
\cos (L-1)k_{1} & \cdots & \cos (L-1)k_{L/2 -1} & -1/\sqrt{2}
\end{array} \right)
\label{umat}
\end{equation}
Here, the wavevectors are $k_j = 2\pi j/L$; note that $U$ is orthogonal
and therefore the Jacobian term in the free energy (\ref{modprin}) gives
no contribution for the Fourier choice of variables. For convenience in
notation, the index $j$ labeling the variables $B_j $ in (\ref{foutra}),
i.e. the columns in (\ref{umat}), will be taken to run
from $-L/2 +1 $ to $L/2$.

In the Fourier basis, the kinetic part of the action (\ref{sq1def}) is
exactly diagonalized and coincides with its free partner
\begin{equation}
\tilde{S}_{Q1}^{kin} (B_i ) =
\frac{2}{N} \mbox{Tr} \, \sum_{j=-L/2 +1}^{L/2} (1-\cos k_j ) B_j^2
=\tilde{S}_{Q1f}^{kin} (B_i )
\end{equation}
On the other hand, the potential part
in the new variables is
\begin{equation}
\tilde{S}_{Q1}^{pot} (B_i ) = \frac{g}{N} \mbox{Tr} \, \sum_{i} \sum_{jklm}
U_{ij} B_j U_{ik} B_k U_{il} B_l U_{im} B_m
\end{equation}
Assuming as before $\langle B_i \rangle =0$, the only terms which
will give a contribution to the free partner are ones in which
the four indices $j,k,l,m$ all take the same value or pairwise two
different values. Thus, one has the free partner
\begin{equation}
\tilde{S}_{Q1f}^{pot} (B_i ) = g\left[
4\sum_{j<k} \frac{1}{N^2 } \mbox{Tr} \, B_j^2 \mbox{Tr} \, B_k^2 
\left( \sum_{i} U_{ij}^2 U_{ik}^2 \right)
+\sum_{j} \frac{1}{N} \mbox{Tr} \, B_j^4 
\left( \sum_{i} U_{ij}^4 \right) \right]
\end{equation}
By explicit calculation, $\sum_{i} U_{ij}^4 $ is of order $O(1/L)$ and
therefore the quartic term can be dropped for a large number of time
slices $L$. On the other hand, again by explicit calculation,
\begin{equation}
\sum_{i=0}^{L-1} U_{ij}^2 U_{ik}^2 = \frac{1}{L} \ \ \ \ \ \ \ \ 
\mbox{for all} \ j,k
\end{equation}
and therefore one finally has for the free partner of the action
(\ref{sq1def}) in the Fourier variables $B_i $,
\begin{equation}
\tilde{S}_{Q1f} (B_i ) = 2 \sum_{i=-L/2 +1}^{L/2} (1-\cos k_i )
\frac{1}{N} \mbox{Tr} \, B_i^2
+\frac{4g}{L} \sum_{i<j=-L/2 +1}^{L/2} 
\frac{1}{N^2 } \mbox{Tr} \, B_i^2 \mbox{Tr} \, B_j^2
\end{equation}
On the other hand, the free partner of (\ref{sq1def}) in the original
variables is
\begin{equation}
S_{Q1f} (M_i ) = \frac{1}{N} \mbox{Tr} \, \sum_{i=0}^{L-1} 
(2M_i^2 + gM_i^4 )
\label{sforig}
\end{equation}
again assuming $\langle M_i \rangle =0$.
Here, the potential term is treated exactly, whereas the nearest-neighbor
coupling has been completely truncated.

The free models are now easily solved. For the purpose of calculating
the free energies, the reference action $V_{ref} (C_i ) = (1/N) \mbox{Tr} \,
\sum_{i} C_i^2 $ will be used; then one considers (cf. Eq. (\ref{dpridef}))
\begin{equation}
S^{\prime }_{Q1f} (M_i ) = \frac{1}{N} \mbox{Tr} \, \sum_{i}
\left[ (1+\alpha ) M_i^2 + \alpha g M_i^4 \right]
\label{sfcorig}
\end{equation}
for the original variables and
\begin{equation}
\tilde{S}^{\prime }_{Q1f} (B_i ) = \sum_{i} (1+\alpha -2\alpha \cos k_i ) 
\frac{1}{N} \mbox{Tr} \, B_i^2 + \frac{4\alpha g}{L} \sum_{i<j}
\frac{1}{N^2 } \mbox{Tr} \, B_i^2 \mbox{Tr} \, B_j^2
\end{equation}
for the Fourier variables. In the latter, introducing the abbreviation
\begin{equation}
x_B =\frac{2}{L} \left\langle \sum_{i} B_i^2 \right\rangle
\end{equation}
one has semicircular distributions for the matrix variables $B_i $
of radius squared
\begin{equation}
r_i^2 = \frac{2}{1+\alpha -2\alpha \cos k_i + \alpha gx_B }
\label{rsbi}
\end{equation}
implying second moments $\langle B_i^2 \rangle =r_i^2 /4 $, which
self-consistently determines $x_B $:
\begin{eqnarray}
x_B = \frac{2}{L} \left\langle \sum_{i} B_i^2 \right\rangle &=&
\frac{1}{L} \sum_{i} \frac{1}{1+\alpha -2\alpha \cos (2\pi i/L)
+\alpha gx_B } \nonumber \\
& \stackrel{L\rightarrow \infty }{\longrightarrow } &
\int_{-1/2}^{1/2} dk \, \frac{1}{1+\alpha +\alpha gx_B  
-2\alpha \cos 2\pi k} \\
&=& \frac{1}{\sqrt{(\alpha +\alpha gx_B +1)^2 - 4\alpha^{2} } }
\nonumber
\end{eqnarray}
or, 
\begin{equation}
x_B^2 (\alpha +\alpha gx_B +1)^2 -4\alpha^{2} x_B^2 -1 =0
\label{xq1eq}
\end{equation}
The free energy is then
\begin{eqnarray}
F_B &=& \int_{0}^{1} d\alpha \, \sum_{i} \frac{1}{2}
\frac{2-2\cos (2\pi i/L) + gx_B -1}{1+\alpha -2\alpha \cos (2\pi i/L)
+\alpha gx_B } \nonumber \\
& \stackrel{L\rightarrow \infty }{\longrightarrow } &
L \int_{0}^{1} d\alpha \, \int_{-1/2}^{1/2} dk \, \frac{1}{2}
\frac{2-2\cos 2\pi k + gx_B -1}{1+\alpha -2\alpha \cos 2\pi k
+\alpha gx_B } \\
&=& L \int_{0}^{1} d\alpha \, \frac{1}{2\alpha } \left[
1-\frac{1}{\sqrt{(\alpha +\alpha gx_B +1)^2 -4\alpha^{2} } } \right]
\nonumber
\end{eqnarray}
which must be calculated numerically (remember that at every $\alpha $,
$x_B $ is determined by the equation (\ref{xq1eq})).

On the other hand, in the decoupled basis, one has the quartic action
(\ref{sfcorig}), identical and independent for all the $M_i $. Using
Appendix \ref{appa}, the resulting moments are
\begin{eqnarray}
\langle M^2 \rangle &=& \frac{1}{108\alpha^{2} g^2 } \left[
-(1+\alpha )^3 - 18\alpha g (1+\alpha )
+((1+\alpha )^2 + 12\alpha g)^{3/2} \right] \\
\langle M^4 \rangle &=& \frac{1}{216\alpha^{3} g^3 } \left[
(1+\alpha )^4 + 18\alpha g (1+\alpha )^2 + 54\alpha^{2} g^2
-(1+\alpha ) ((1+\alpha )^2 + 12\alpha g)^{3/2} \right]
\end{eqnarray}
and, inserting in the free energy,
\begin{eqnarray}
F_M &=& \int_{0}^{1} d\alpha \, 2L \langle M^2 \rangle +
Lg \langle M^4 \rangle - L \langle M^2 \rangle \nonumber \\
&=& \frac{L}{108 g^2 } \left[ -4-36g-\frac{81g^2 }{2} +
(4+30g)\sqrt{1+3g} + 54g^2 \ln (1+\sqrt{1+3g} ) \right]
\end{eqnarray}
Numerically, $F_M = F_B $ at $g=1.2306$. For smaller $g$, the Fourier
basis provides the better approximation, for larger $g$ the original
choice of variables does. It has been argued in \cite{GroK}
that matrix models such as the one considered here exhibit a
Kosterlitz-Thouless transition at some value of the lattice
spacing separating the two regimes where the matrix chain
behaves more like a collection of independent sites and the one
where it supports spin wave type of modes. The variational approach
presented here, while giving no details of the transition region,
is capable of capturing these two different possible regimes.

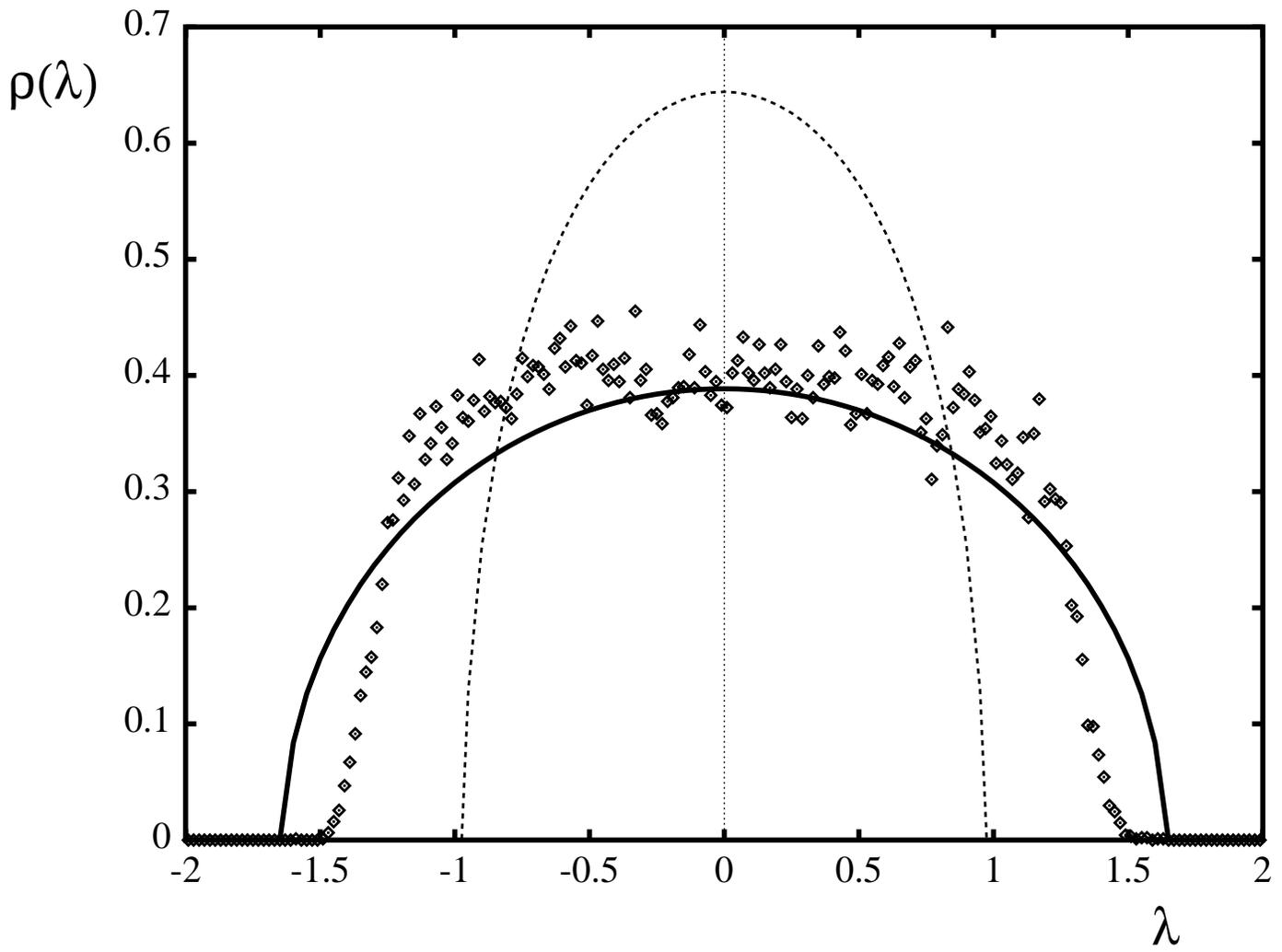
\begin{figure}
\caption{Eigenvalue distribution $\rho (\lambda )$ 
generated by $S_{Q1} $ at $g=0.1$.
Data points: Monte Carlo result for $N=10$ and $L=10$. Solid line: Free
variational approximation in the Fourier basis. Dotted: Free
variational approximation in the original basis. The former
is the better approximation of the two.}
\label{fig9}
\end{figure}

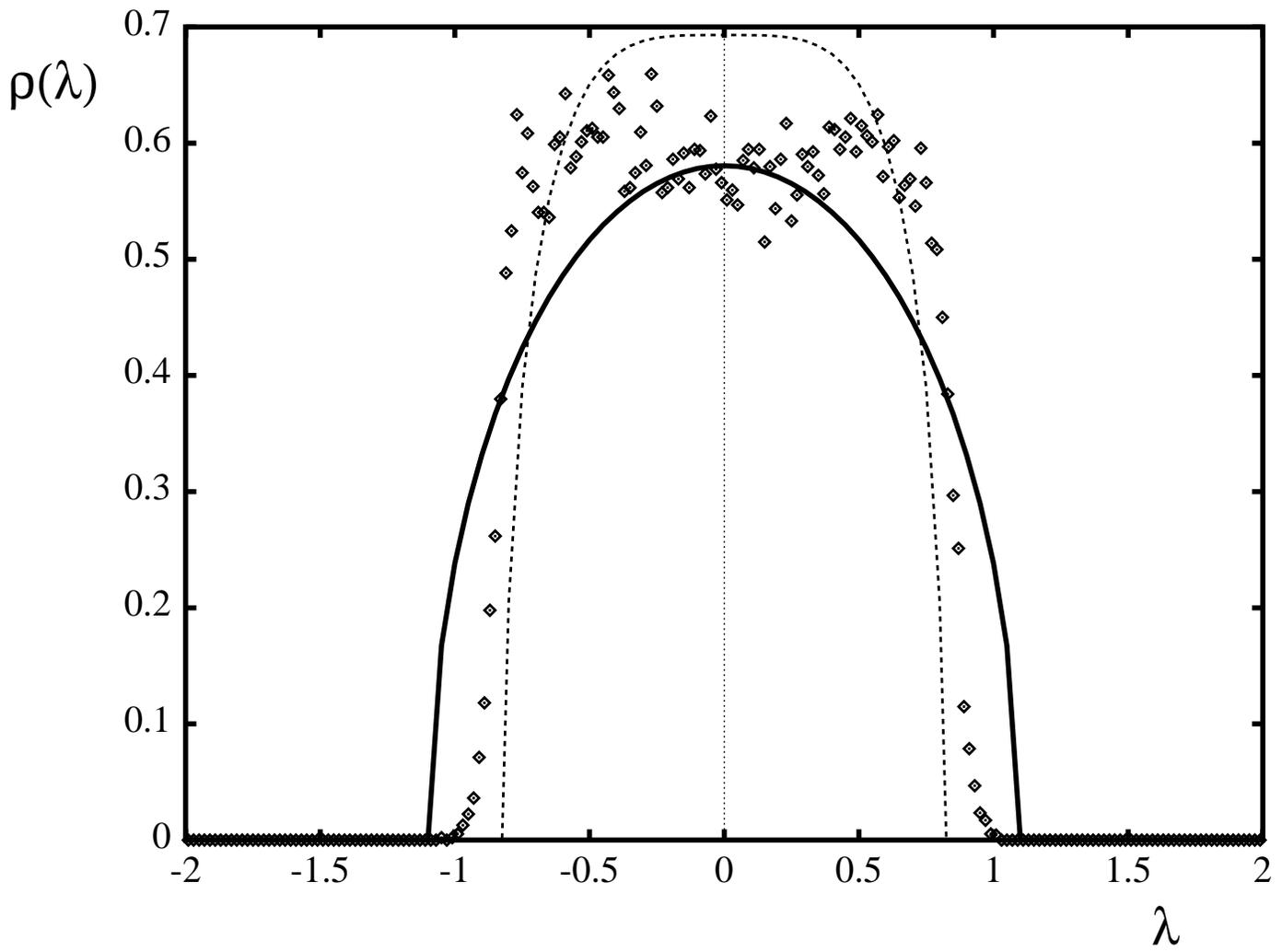
\begin{figure}
\caption{Eigenvalue distribution $\rho (\lambda )$ 
generated by $S_{Q1} $ at $g=1$.
Data points: Monte Carlo result for $N=10$ and $L=10$. Solid line: Free
variational approximation in the Fourier basis. Dotted: Free
variational approximation in the original basis. The former
is the better approximation of the two.}
\label{fig10}
\end{figure}

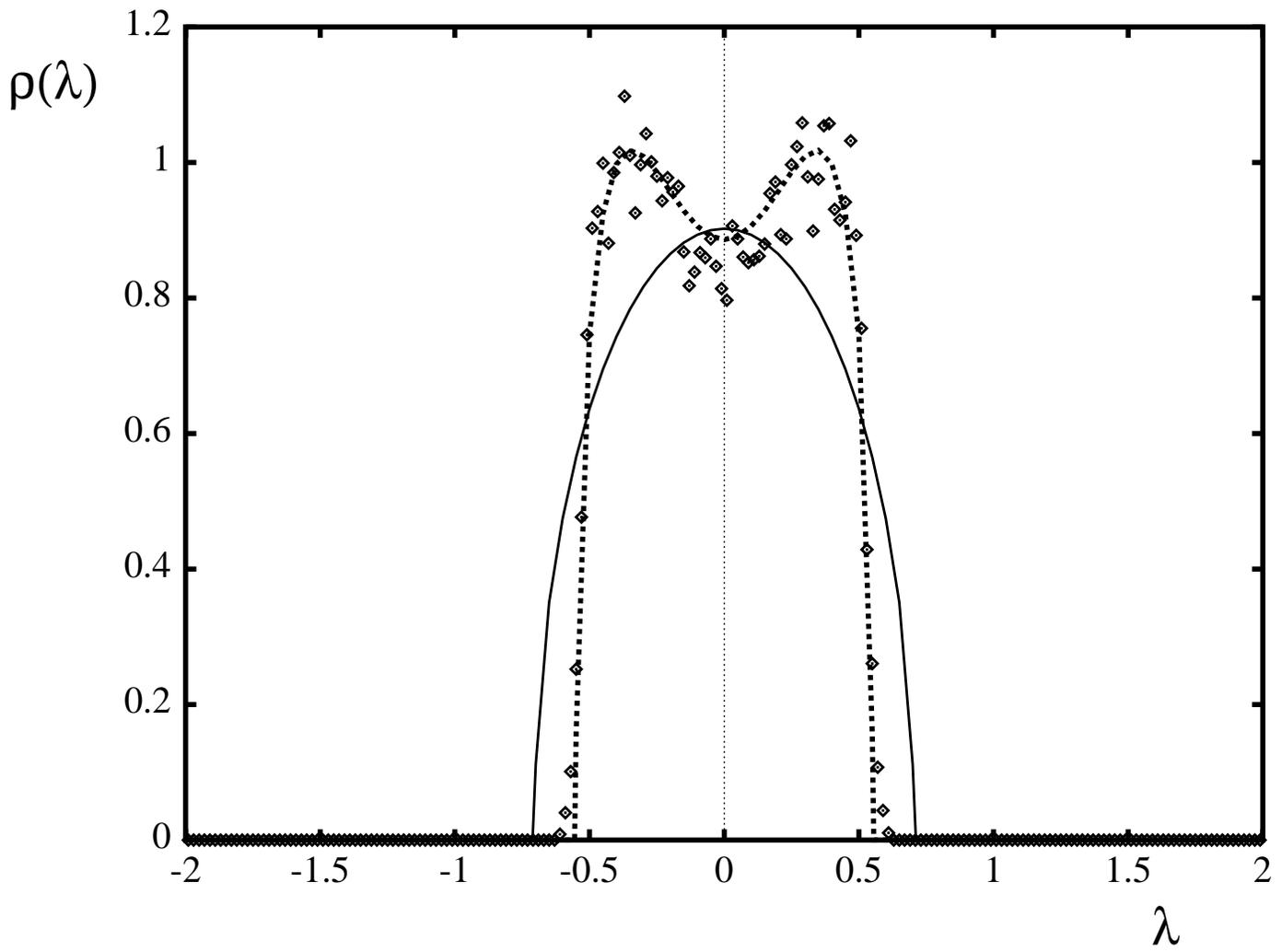
\begin{figure}
\caption{Eigenvalue distribution $\rho (\lambda )$ 
generated by $S_{Q1} $ at $g=10$.
Data points: Monte Carlo result for $N=10$ and $L=10$. Solid line: Free
variational approximation in the Fourier basis. Dotted: Free
variational approximation in the original basis. The latter
is the better approximation of the two.}
\label{fig11}
\end{figure}

It remains to give the eigenvalue distributions in the two regimes. In
the original variables, described by the action $S_{Q1f} (M_i )$, cf.
equation (\ref{sforig}), one immediately has, using Appendix \ref{appa},
\begin{equation}
\rho_{M} (\lambda ) = \frac{1}{\pi } (2g\lambda^{2} + 2 + m^2 g)
\sqrt{m^2 -\lambda^{2} } \ \ \ \ \ \ \mbox{with} \ \ 
m^2 = \frac{2}{3g} (\sqrt{1+3g} -1)
\end{equation}
for all the variables $M_i $. In the Fourier case, one must convolute
the semicircular distributions of the $B_i $ to obtain the distributions
of the $M_i $. It is well known \cite{Voi} that the semicircular
distributions are closed under additive convolution and that the
square radii add up to the square radius of the resulting semicircle.
I.e.,
\begin{equation}
r_{M_i }^{2} = \sum_{j} U_{ij}^{2} r_{B_j }^{2}
\end{equation}
Inserting (\ref{rsbi}) for $\alpha =1$ and (\ref{umat}), one obtains
identical radii for all the $M_i $,
\begin{eqnarray}
r_{M_i }^{2} &=& \frac{1}{L} \sum_{j=-L/2 +1}^{L/2}
\frac{2}{2-2\cos (2\pi j/L) + gx_B } \nonumber \\
& \stackrel{L\rightarrow \infty }{\longrightarrow } &
\int_{-1/2}^{1/2} dk \,
\frac{2}{2-2\cos 2\pi k +gx_B } \\
&=& \frac{2}{\sqrt{4gx_B + g^2 x_B^2 } } \nonumber \\
&=& 2x_B \nonumber
\end{eqnarray}
where in the last line it has been used that 
$x_B =(2/L) \langle \sum_{i} B_i^2 \rangle $
solves (cf. (\ref{xq1eq}) for $\alpha =1$)
\begin{equation}
4gx_B^3 + g^2 x_B^4 =1
\label{q1bx}
\end{equation}
The corresponding semicircular distribution for the $M_i $ is
then
\begin{equation}
\rho_{M} (\lambda ) = \frac{1}{x_B \pi } \sqrt{2x_B -\lambda^{2} }
\label{q1brho}
\end{equation}
The variational approximation derived above should now be compared to
Monte Carlo experiments and the exact results known about this model.
In Figs. \ref{fig9}-\ref{fig11} Monte Carlo results 
for $L=10$ and $N=10 $ are
exhibited together with variational results at different couplings $g$.
Checks with $N=20$ and $L=20$ revealed no significant differences.
One indeed observes, as already borne out by the calculation of the
free energies, that the Fourier basis is preferable in a weak
potential and the original basis is favored by a strong potential.
This is corroborated by the Monte Carlo evaluation of correlators
between matrices at neighboring time slices, cf. Table 4. 

\begin{center}
\begin{tabular}{|c||c|c||c|c||c|c||c|c|}
\hline \rule[-1.5ex]{0ex}{4.5ex} $g$ &
$\langle M_i^2 \rangle $ & $\langle M_i^4 \rangle $ &
$C_1 $ & $C_{1f} $ & $C_2 $ & $C_{2f} $ & $C_3 $ & $C_{3f} $ \\
\hline \hline
.1 &
.55 & .57 & .41 & .31 & .59 & .33 & .52 & .26 \\ \hline
1 &
.24 & .11 & .066 & .059 & .015 & .012 & .014 & .0093 \\ \hline
10 &
.098 & .017 & .0099 & .0097 & .00031 & .00029 & .00025 & .00024 \\ \hline
\end{tabular} \vspace{0.3cm}

Table 4 : Mixed correlators, Eqs. (\ref{mixmf}), between matrices $M_i $ 
at neighboring
time slices and predictions for free $M_i $
\end{center}

At large coupling, neighboring matrices are to a very good approximation
free, meaning the original decoupled basis gives a good description.
By contrast, at low $g$, there are significant angular correlations
induced by the nearest-neighbor coupling. These angular correlations are
washed out by the fluctuations when one considers matrices far apart
on the temporal lattice, cf. Table 5. There is virtually no deviation
from free behavior between distant matrices even at low values of $g$.

\begin{center}
\begin{tabular}{|c||c|c||c|c||c|c||c|c|}
\hline \rule[-1.5ex]{0ex}{4.5ex} $g$ &
$\langle M_i^2 \rangle $ & $\langle M_i^4 \rangle $ &
$C_1 $ & $C_{1f} $ & $C_2 $ & $C_{2f} $ & $C_3 $ & $C_{3f} $ \\
\hline \hline
.1 &
.55 & .57 & .31 & .31 & .35 & .33 & .28 & .26 \\ \hline
1 &
.24 & .11 & .059 & .059 & .012 & .012 & .0093 & .0093 \\ \hline
10 &
.098 & .017 & .0097 & .0097 & .00029 & .00029 & .00024 & .00024 \\ \hline
\end{tabular} \vspace{0.3cm}

Table 5 : Mixed correlators, Eqs. (\ref{mixmf}), between matrices $M_i $ 
separated by half of
the total length of the temporal lattice, and predictions for free $M_i $
\end{center}

In order to make contact with physical (still Euclidean) time,
one should write the discretized action of the quantum one-matrix
model as
\begin{equation}
S_{Q1}^{\prime } = \frac{1}{N} \mbox{Tr} \, \left( \sum_{i=0}^{L-1} 
\frac{(M_{i+1}^{\prime } -M_i^{\prime } )^2 }{2} \frac{L}{T}
+ \tilde{g} \frac{T}{L} (M_i^{\prime } )^4 \right)
\end{equation}
where $T$ is now the length of the circle in time direction.
Rescaling $M_i^{\prime } = M_i \sqrt{2T/L} $, one regains the
form (\ref{sq1def}) of the action, with the identification
$g=4\tilde{g} T^3 /L^3 $. Therefore, $g$ behaves like the third
power of the lattice spacing.

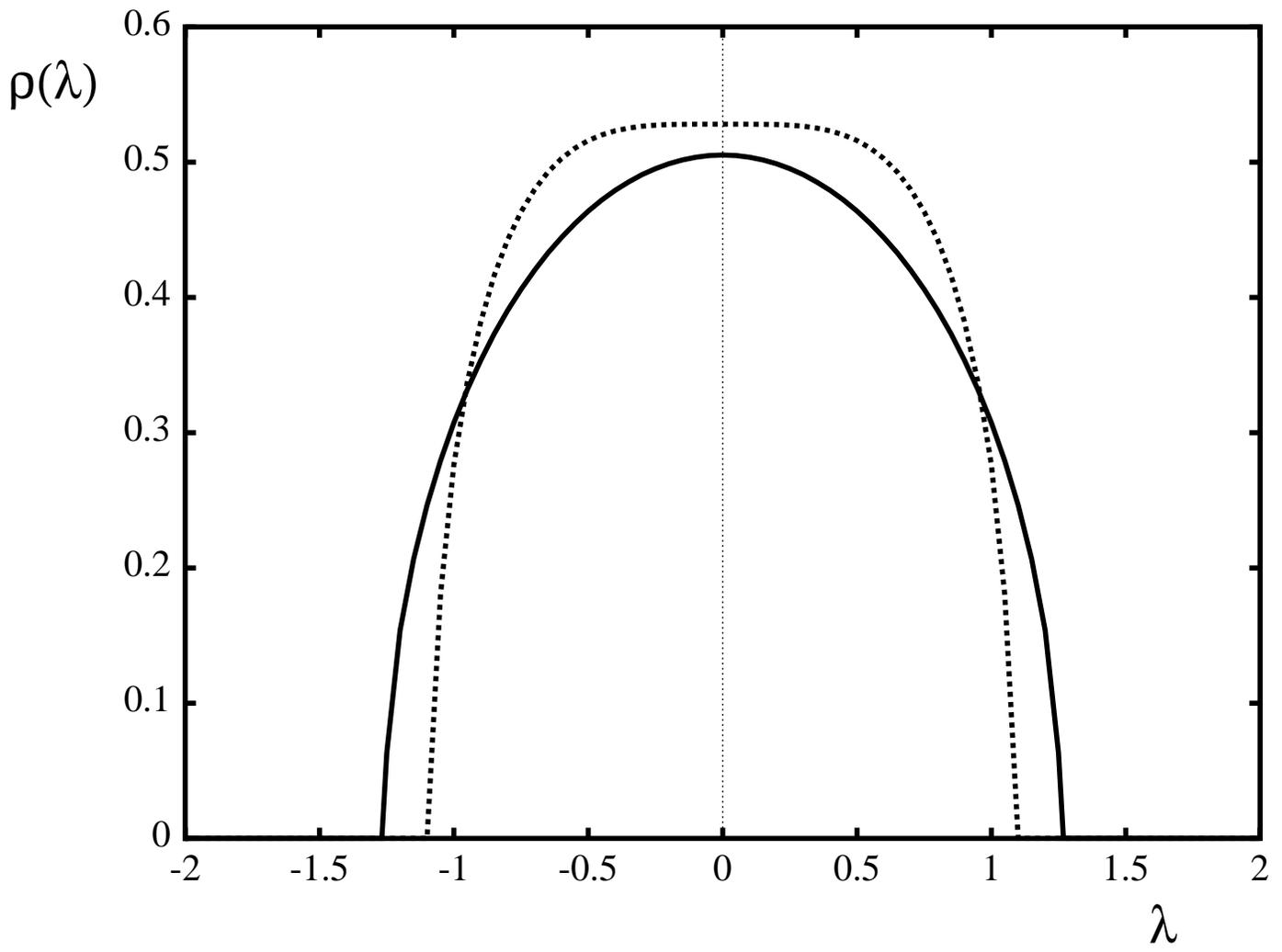
\begin{figure}
\caption{Eigenvalue distribution $\rho (\lambda )$ 
generated by $S_{Q1} $ in the continuum limit,
namely for $\tilde{g} =1$ (for the definition
of $\tilde{g} $, see text). Dotted: Exact result. Solid line:
Free variational approximation in the Fourier basis.}
\label{fig12}
\end{figure}

In the continuum limit, the ground state distribution 
of the eigenvalues of the quantum mechanical
one-matrix model has been found analytically in \cite{Bre1}.
This can now be compared to the variational solution. According
to the correspondence established above, the limit of vanishing
lattice spacing corresponds to vanishing $g$; therefore the
Fourier basis is the preferable one. The eigenvalue distribution
for the rescaled variables $M_i^{\prime } $ is 
(cf. equation (\ref{q1brho}))
\begin{equation}
\rho_{M^{\prime } } (\lambda ) = \sqrt{\frac{L}{2T} } 
\rho_{M} \left( \sqrt{\frac{L}{2T} } \lambda \right) =
\sqrt{\frac{L}{2T} } \frac{1}{x_B \pi } 
\sqrt{2x_B -\frac{L}{2T} \lambda^{2} }
\end{equation}
and $x_B $ is determined by (cf. equation (\ref{q1bx}))
\begin{equation}
16\tilde{g} \frac{T^3 }{L^3 } x_B^3 + 16\tilde{g}^{2} \frac{T^6 }{L^6 }
x_B^4 =1
\end{equation}
which for large $L$ is solved by
\begin{equation}
x_B =\frac{L}{T} (16\tilde{g} )^{-1/3} + O(1)
\end{equation}
Therefore
\begin{equation}
\rho_{M^{\prime } } (\lambda ) = \frac{2}{\pi }
\left( \frac{\tilde{g} }{4} \right)^{1/3}
\sqrt{ \left( \frac{4}{\tilde{g} } \right)^{1/3} -\lambda^{2} }
\end{equation}
Note that there is no $T$-dependence left in the continuum limit.
By contrast, the exact solution is \cite{Bre1}
\begin{equation}
\rho_{M^{\prime } }^{exact} (\lambda ) = \frac{1}{\pi }
\sqrt{2\epsilon -2\tilde{g} \lambda^{4} } \ \ \ \ \ \ \ 
\mbox{with} \ \ \ \ \epsilon^{3} = 81\pi^{6} \tilde{g} 
\left( \frac{1}{\Gamma (1/4) } \right)^{8}
\end{equation}
determined by the normalization condition. The two solutions are
compared for $\tilde{g} =1$ in Fig. \ref{fig12}.

\subsection{Quantum Mechanics of Two Matrices}
While the case of one quantum mechanical matrix discussed in the
previous section is amenable to exact treatment \cite{Bre1} due to
its special form, an analogous solution is not possible once more than 
one quantum mechanical matrix is involved. On the other hand, the
approximate variational approach, using discretized time, is easily
generalized to the case of more than one matrix. Here, a commutator-type
interaction will be considered, as it is especially interesting from the
point of view of applications to Yang-Mills theories:
\begin{equation}
S_{Q2} = \frac{1}{N} \mbox{Tr} \, \left( \sum_{i=0}^{L-1} 
(M_{1,i+1} -M_{1,i} )^2 + (M_{2,i+1} -M_{2,i} )^2
+ g(i [ M_{1,i},M_{2,i} ] )^2 \right)
\label{sq2def}
\end{equation}
where the first index labels the two different matrix variables and
the second one the different time slices. Again, only the original 
choice of variables will be compared to a Fourier basis (cf. (\ref{umat}))
using the variational criterion.

Note that the action $S_{Q2} $ is invariant under simultaneous shifts
of all the matrices by a multiple of the unit matrix. This trivial
freedom should be removed in order to obtain stable Monte Carlo
results; otherwise, the system just performs a random walk in the
trace of the matrices. Thus, the $M_{n,i} $ here are constrained to
be traceless, i.e. the condition $\langle M_{n,i} \rangle =0$,
assumed to be realized dynamically in prior examples, is enforced by
hand.

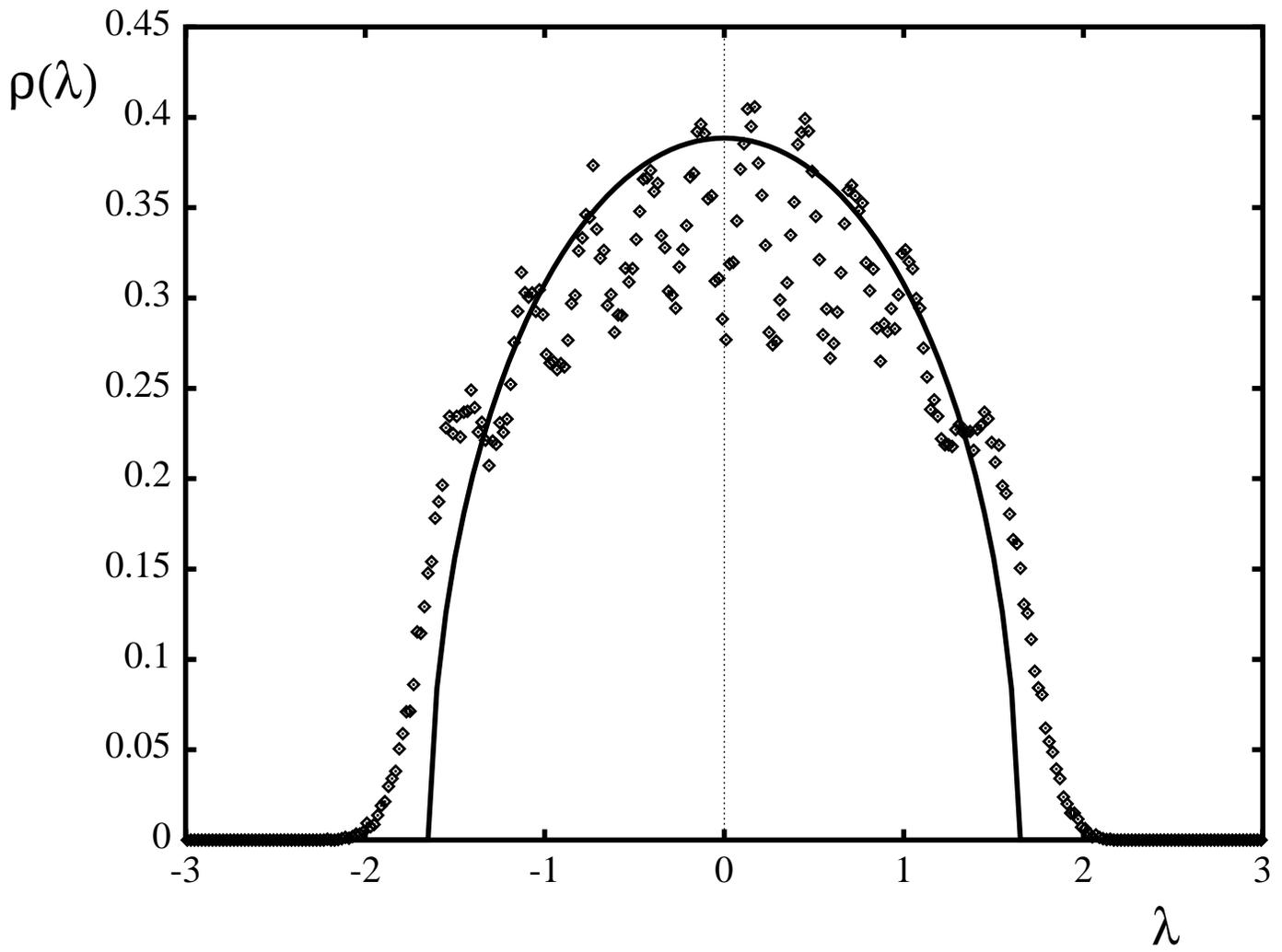
\begin{figure}
\caption{Eigenvalue distribution $\rho (\lambda )$ 
generated by $S_{Q2} $ at $g=0.1$.
Data points: Monte Carlo result for $N=10$ and $L=10$. Solid line:
Free variational approximation in the Fourier basis.}
\label{fig13}
\end{figure}

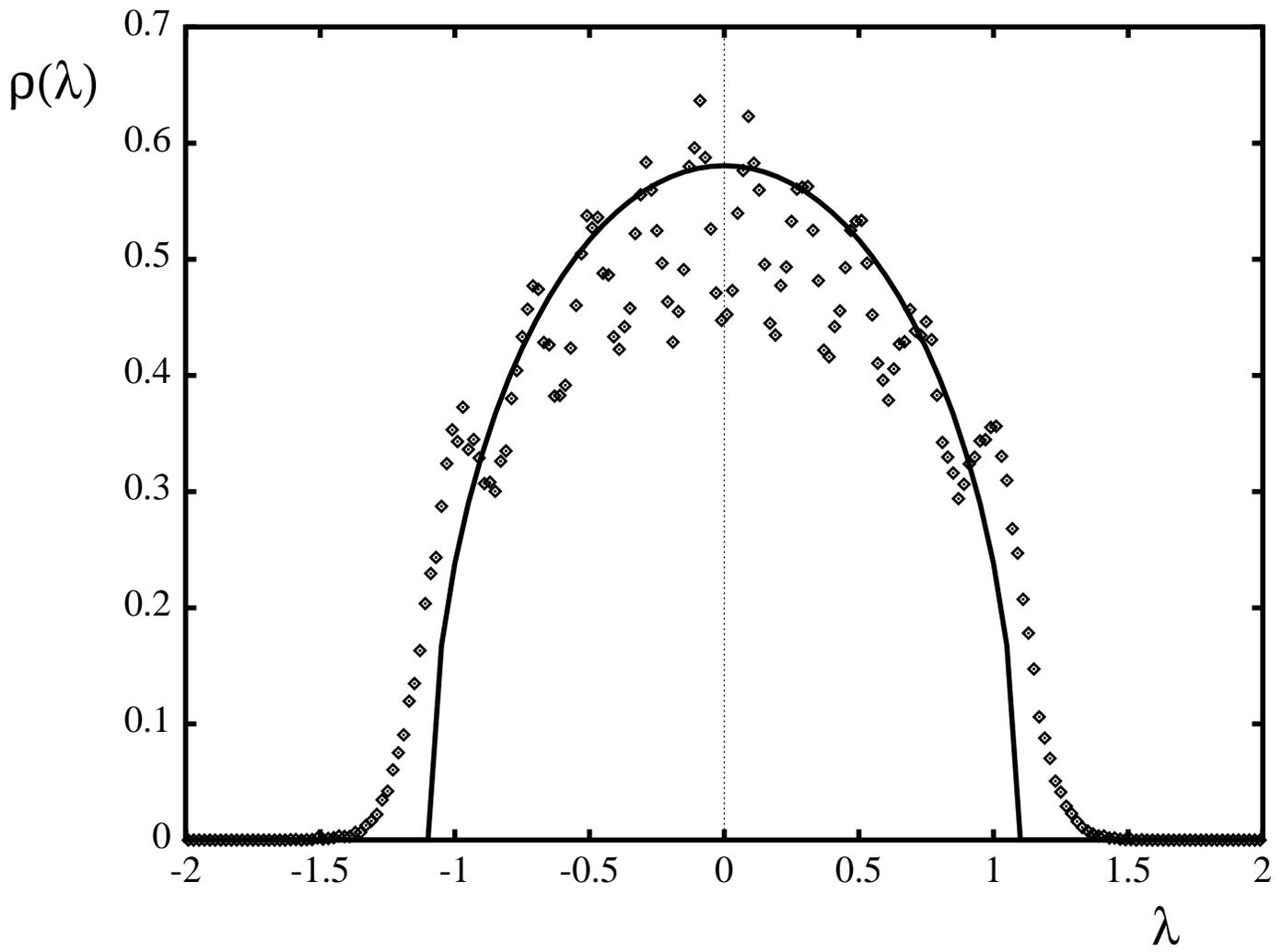
\begin{figure}
\caption{Eigenvalue distribution $\rho (\lambda )$ 
generated by $S_{Q2} $ at $g=1$.
Data points: Monte Carlo result for $N=10$ and $L=10$. Solid line: 
Free variational approximation in the Fourier basis.}
\label{fig14}
\end{figure}

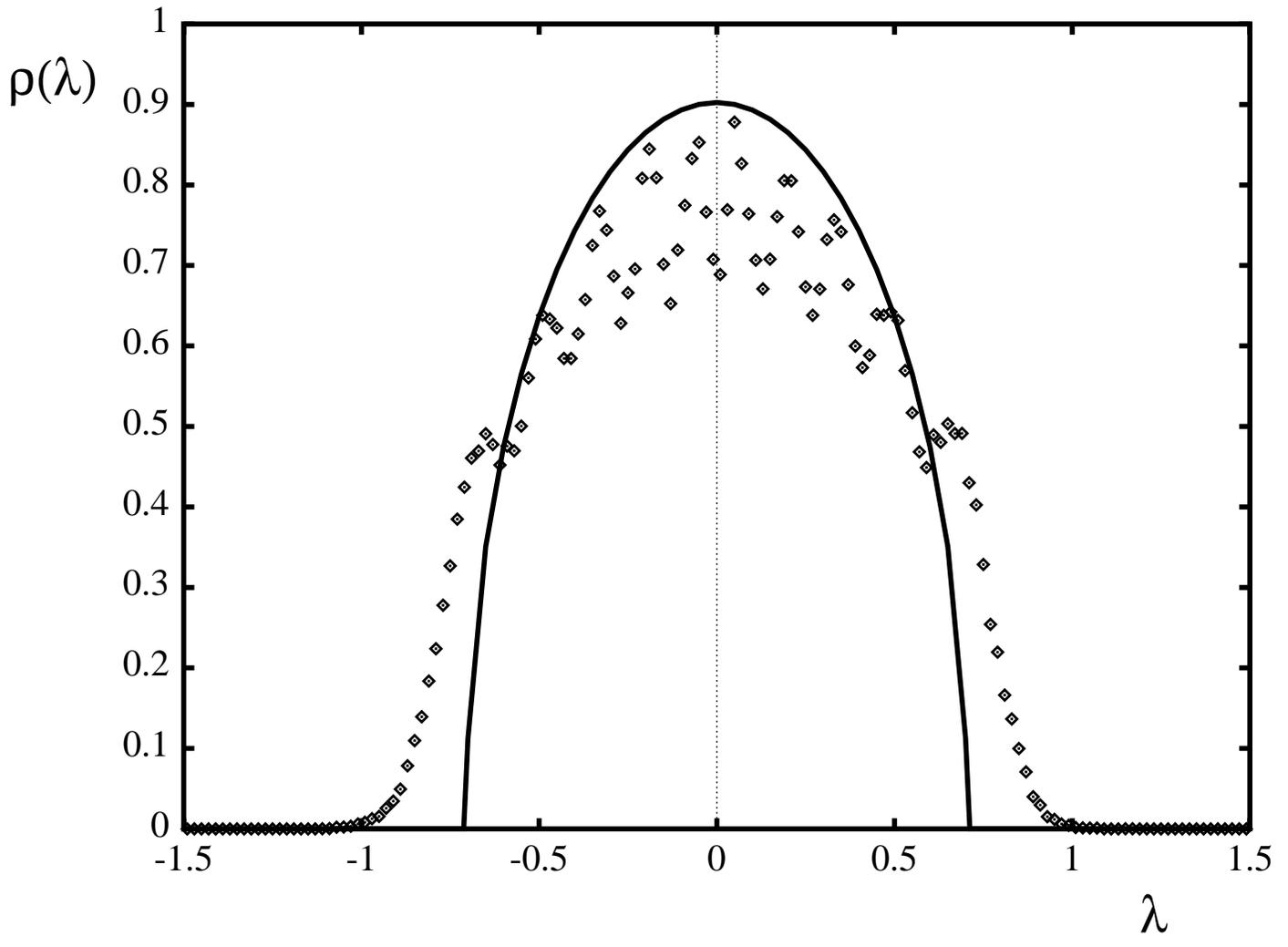
\begin{figure}
\caption{Eigenvalue distribution $\rho (\lambda )$ 
generated by $S_{Q2} $ at $g=10$.
Data points: Monte Carlo result for $N=10$ and $L=10$. Solid line: 
Free variational approximation in the Fourier basis.}
\label{fig15}
\end{figure}

The free partner to (\ref{sq2def}) in the original basis is
\begin{equation}
S_{Q2f} (M_{n,i} ) = \sum_{i=0}^{L-1} \frac{1}{N} \mbox{Tr} \,
(2M_{1,i}^2 + 2M_{2,i}^2 )
+ \frac{2g}{N^2 } \mbox{Tr} \, M_{1,i}^2 \mbox{Tr} \, M_{2,i}^2
\label{sq2fdef}
\end{equation}
(using $\langle M_{1,i} \rangle = \langle M_{2,i} \rangle =0$).
On the other hand, in the Fourier basis
\begin{equation}
M_{n,i} = \sum_{j} U_{ij} B_{n,j}
\end{equation}
with $U$ as in (\ref{umat}), the kinetic part becomes
\begin{equation}
\tilde{S}_{Q2}^{kin} (B_{n,i} ) = \tilde{S}_{Q2f}^{kin} (B_{n,i} ) =
\frac{2}{N} \mbox{Tr} \, \sum_{j=-L/2 +1}^{L/2} (1-\cos k_j ) 
(B_{1,j}^2 + B_{2,j}^2 )
\end{equation}
The potential part on the other hand becomes
\begin{equation}
-\frac{g}{N} \mbox{Tr} \, \sum_{i} [ M_{1,i},M_{2,i} ]^2 =
\frac{2g}{N} \mbox{Tr} \, \sum_{i} \sum_{jklm}
U_{ij} U_{ik} U_{il} U_{im}
(B_{1,j} B_{1,k} B_{2,l} B_{2,m} - B_{1,j} B_{2,k} B_{1,l} B_{2,m} )
\end{equation}
Using again the assumption $\langle B_{n,i} \rangle =0$, the second term
in the round brackets on the right hand side never gives a contribution
to the free partner, whereas the first term only gives a contribution
if $j=k$ and $l=m$. Thus one has
\begin{equation}
\tilde{S}_{Q2f}^{pot} (B_{n,i} ) =
\frac{2g}{N^2 } \sum_{j,k} \mbox{Tr} \, B_{1,j}^2 \mbox{Tr} \, B_{2,k}^2
\sum_{i} U_{ij} U_{ij} U_{ik} U_{ik}
=\frac{2g}{N^2 L} \sum_{j,k} 
\mbox{Tr} \, B_{1,j}^2 \mbox{Tr} \,  B_{2,k}^2
\end{equation}
Now one can again easily solve the free models. For the purpose of 
calculating the free energies, the reference action
\begin{equation}
V_{ref} (C_{n,i} ) = 
\frac{1}{N} \mbox{Tr} \, \sum_{i} (C_{1,i}^2 + C_{2,i}^2 ) 
\end{equation}
is convenient; then one has to solve the models
\begin{equation}
S^{\prime }_{Q2f} (M_{n,i} ) = 
\sum_{i=0}^{L-1} \frac{1+\alpha }{N} \mbox{Tr} \, 
(M_{1,i}^2 + M_{2,i}^2 ) + \frac{2\alpha g}{N^2 }
\mbox{Tr} \, M_{1,i}^2 \mbox{Tr} \, M_{2,i}^2
\label{q2fcor}
\end{equation}
for the original variables $M_{n,i} $ and
\begin{equation}
\tilde{S}^{\prime }_{Q2f} (B_{n,i} ) = \sum_{j} (1+\alpha -2\alpha \cos k_j )
\frac{1}{N} \mbox{Tr} \, (B_{1,j}^2 + B_{2,j}^2 )
+\frac{2g\alpha}{N^2 L} \sum_{j,k} 
\mbox{Tr} \, B_{1,j}^2 \mbox{Tr} \, B_{2,k}^2
\label{q2fcfo}
\end{equation}
for the Fourier variables $B_{n,i} $.
Starting with (\ref{q2fcor}), abbreviating 
$\langle M_{n,i}^{2} \rangle =x_M $, one obtains identical semicircular
distributions for all matrix variables with radius squared
\begin{equation}
r^2 = \frac{2}{\alpha +2\alpha gx_M +1}
\end{equation}
whereupon the consistency condition determining $x_M $ becomes
\begin{equation}
x_M =\frac{r^2 }{4} = \frac{1}{2} \frac{1}{\alpha +2\alpha gx_M +1}
\end{equation}
solved by 
\begin{equation}
x_M =\frac{1}{4\alpha g} (\sqrt{(1+\alpha )^2 +4g\alpha } -(1+\alpha ))
\end{equation}
The free energy then is given by
\begin{equation}
F_M = L \int_{0}^{1} d\alpha \, ( 4x_M + 2gx_M^2 - 2x_M )
\end{equation}
On the other hand, in the Fourier case (\ref{q2fcfo}), abbreviating
\begin{equation}
\left\langle \frac{2}{L} \sum_{j} B_{1,j}^2 \right\rangle =
\left\langle \frac{2}{L} \sum_{j} B_{2,j}^2 \right\rangle = x_B
\end{equation}
one notices that the variables $B_{1,j} $ and $B_{2,j} $ are controlled
by exactly the same potential as the variables $B_j $ in the case of the
quantum mechanical one-matrix model discussed in the previous section.
In particular, one has the consistency condition
\begin{equation}
x_B^2 (\alpha +\alpha gx_B +1)^2 -4\alpha^{2} x_B^2 -1 =0
\end{equation}
and the second moments
\begin{equation}
\langle B_{1,j}^2 \rangle = \langle B_{2,j}^2 \rangle =
\frac{1}{2} \frac{1}{1+\alpha -2\alpha \cos k_j + \alpha gx_B }
\end{equation}
The free energy now is
\begin{equation}
F_B = L \int_{0}^{1} d\alpha \, \left[ 2x_B - \frac{1}{L}
\sum_{j=-L/2 +1}^{L/2}
\frac{2\cos (2\pi j/L)}{1+\alpha -2\alpha \cos (2\pi j/L) +\alpha gx_B }
+\frac{gx_B^2 }{2} -x_B \right]
\end{equation}
Numerical evaluation of the two free energies yields $F_B \leq F_M $
for all $g$, i.e. the Fourier basis is always preferable over the
original basis. While this is to be expected at small $g$, it is not
necessarily an indication that the Fourier basis represents
a particularly good approximation for large $g$ as well: Rather,
working in the original basis still truncates the interaction term
rather badly at every time slice (cf. equation (\ref{sq2fdef})).
This is different from
the one-matrix case, where assuming the original variables to be free merely
implied decoupling variables at different time slices, which certainly
should become exact for large coupling $g$. In the two-matrix case,
by contrast, there is still an additional truncation at every time
slice which prevents convergence to the exact
problem at large $g$. In light of this, it is not so surprising that the
Fourier basis is preferable for all $g$. Nevertheless, in Figs.
\ref{fig13}-\ref{fig15}, this basis appears to provide quite a
satisfactory description even for $g=10$; the agreement with Monte
Carlo results ($N=10$ and $L=10$ were used in the latter) is not much
worse at $g=10$ than at $g=1/10$.

Also in the correlators, cf. Tables 6 and 7, one observes that the
angular correlations involving the same matrix variable at different
time slices are stronger than the angular correlations between the
two different matrix variables at a fixed time, up to quite high $g$.
At $g=10$, these two correlations seem to be roughly equally strong.
This again corroborates the result that using the Fourier basis,
which treats the nearest-neighbor coupling exactly, provides a good
approximation for a large range of $g$.

\begin{center}
\begin{tabular}{|c||c|c||c|c||c|c||c|c|}
\hline \rule[-1.5ex]{0ex}{4.5ex} $g$ &
$\langle M_{n,i}^{2} \rangle $ & $\langle M_{n,i}^{4} \rangle $ &
$C_1 $ & $C_{1f} $ & $C_2 $ & $C_{2f} $ & $C_3 $ & $C_{3f} $ \\
\hline \hline
.1 &
.86 & 1.5 & 1.2 & .75 & 5.6 & 2.3 & 5.1 & 1.7 \\ \hline
1 &
.38 & .29 & .19 & .14 & .16 & .084 & .13 & .063 \\ \hline
10 &
.18 & .068 & .038 & .033 & .0064 & .0047 & .005 & .0034 \\ \hline
\end{tabular} \vspace{0.3cm}

Table 6 : Mixed correlators, Eqs. (\ref{mixmf}), between matrices 
$M_{n,i} $ for one fixed
$n$ at neighboring time slices $i$ and predictions for free $M_{n,i} $
\end{center}

\begin{center}
\begin{tabular}{|c||c|c||c|c||c|c||c|c|}
\hline \rule[-1.5ex]{0ex}{4.5ex} $g$ &
$\langle M_{n,i}^{2} \rangle $ & $\langle M_{n,i}^{4} \rangle $ &
$C_1 $ & $C_{1f} $ & $C_2 $ & $C_{2f} $ & $C_3 $ & $C_{3f} $ \\
\hline \hline
.1 &
.86 & 1.5 & .66 & .75 & 1.7 & 2.3 & 1.3 & 1.7 \\ \hline
1 &
.38 & .29 & .12 & .14 & .059 & .084 & .043 & .063 \\ \hline
10 &
.18 & .068 & .027 & .033 & .0027 & .0047 & .002 & .0034 \\ \hline
\end{tabular} \vspace{0.3cm}

Table 7 : Mixed correlators between the matrices $M_{1,i} $ 
and $M_{2,i} $ at a fixed time slice $i$ and predictions for free
$M_{1,i} $, $M_{2,i} $.
\end{center}

Ultimately, though, one would expect a local treatment, which assumes
different time slices to be free with respect to one another, to
provide better agreement at high $g$ if it manages to well
approximate the commutator interaction. This is precisely what is not
achieved if the original variables are assumed to be free. On the
contrary, a nontrivial mixing of the variables $M_{1,i} $ and $M_{2,i} $
at every time slice is necessary such 
as to better accomodate the commutator-type interaction in (\ref{sq2def}).
Note that linear combinations of $M_{1,i} $ and $M_{2,i} $ will not
effect any improvement, since the commutator is invariant under such
linear transformations (up to trivial rescalings of the variables).
It is disappointing that precisely this phenomenologically very interesting
type of interaction term seems particularly resistent to free
approximation combined with linear transformations. Here, it seems
that using (technically more complicated) nonlinear variable
transformations must be contemplated in order to capture the essential
angular correlations. Note also that the commutator term in the action
on its own does not confine the eigenvalue distributions of the
involved variables to a compact support, even if one enforces
tracelessness; the eigenvalues can become arbitrarily large if the
matrices are located in the regions of configuration space where they
commute. This pathology however seems to disappear (according to
Monte Carlo experiments carried out by the authors) when more than
two matrices are involved (i.e. a higher-dimensional Yang-Mills type
of action). Presumably there, the regions of configuration space where
all the matrices commute are too small compared with the whole space
such that entropy suppresses configurations with arbitrarily large 
eigenvalues.

\section{Summary}
In this work, a new variational approach to interacting large-$N$
multi-matrix models was developed. 
The partition function was approximated using the variational
principle $F\le F_0 + \langle S - S_0 \rangle_{S_0 }$ with $S_0 $ 
initially taken from the space of all matrix models which are free in the
set of variables the original interacting problem is given in.
It turned out to be possible to give a general
solution to this variational problem for a fixed set of matrix variables
in terms of the concept of the ``free partner'' introduced by
the authors. The free partner defines a type of mean field
approximation to the original action and is constructed using
the axioms of freeness in a fashion analogous to the construction
of more conventional mean field theories for fermionic or bosonic
particles using Wick's theorem. The freeness axioms constitute the
analog of Wick's theorem for objects obeying the Cuntz algebra,
which is the algebra obeyed by the Fock space operator representation
of free random variables \cite{Gop}.

However, the variational approach presented here goes beyond simply
acting as a device to derive a mean field theory. Above, the variational
space was characterized as being the space of all matrix models which
are free in the set of variables the original interacting problem is
given in. This implies that one can considerably enlarge the variational
space by first allowing for a change of variables in the original problem
and only then varying over all matrix models which are free in the
variables one has settled for. In this way one includes into the variation
models which are not free in the original variables. Every set of 
variables defines a different mean field theory given by the corresponding
free partner. The variational principle not only allows to derive these
mean field theories, but further allows to decide which out of a set of
mean field theories provides the best approximation to the exact problem.

A number of examples were considered in order to assess the accuracy
of the variational method;
the variational results were compared with
Monte Carlo simulations as well as analytical results, where available.
To begin with, classical two-matrix models described by actions with at 
most quartic terms were investigated allowing for a general linear
transformation of matrix variables. Impressive agreement was observed, 
except in the case of the action $S_{1}^{red} $, Eq. (\ref{s1redef}).
The authors did not attempt to improve the approximation by allowing
for nonlinear variable transformations, which may relieve the discrepancy.
Classical models with a large number of matrices
were considered in which all matrices pairwise interact in the
same way. Such models e.g. become relevant when one has managed to
decouple the different space-time points of a large-$N$ matrix field
theory in a large number of dimensions $D$, e.g. in the framework
of an Eguchi-Kawai reduction. Among the models considered was one 
which reduces to $S_{1}^{red} $ for $D=2$. It turned out that the
quality of the variational approximation using only the original set
of variables improves as $D$ rises.

The variational approximation was also considered
for the problems of a single and two coupled matrix chains
simulating path integrals for one and two quantum matrices. Two
discrete choices of free variables, namely the original
variables and a Fourier basis, were considered. In the single chain case,
as expected, for small 
(large) lattice spacing, the Fourier (original) basis provided the
better approximation. In the continuum limit of the one-matrix case,
good agreement was achieved with the available analytical solution.
For two matrix chains coupled via a commutator-type
interaction, the Fourier basis provided a better
approximation for all couplings; it also
compares quite favorably with the exact Monte Carlo results
for a large range of coupling constants. However, for very large
coupling, one would expect a local basis, constructed such as to capture
the angular correlations introduced by the commutator interaction,
to ultimately be more appropriate. Here, again it seemed that technically
more involved nonlinear transformations must be contemplated in order
to achieve further progress.

\appendix

\section{Solution of the one-matrix model in a symmetric quartic
potential}
\label{appa} 
Following \cite{Bre1}, \cite{AkG} 
the one-cut eigenvalue distribution $\rho $ resulting from a hermitian
large-$N$ one-matrix model in a quartic potential $V(x)=a_4 x^4 +
a_2 x^2 $ is most easily found by considering the resolvent, 
\begin{equation} 
G(z):=\int_{-\infty }^{\infty } d\lambda \, 
\frac{\rho  (\lambda  ) }{z-\lambda }  
=\frac{1}{2}  V^{\prime } (z) - Q(z) \sqrt{z^2
-m^2 } 
\end{equation} 
where $Q(z)$ is a polynomial. $Q(z)$, along with $m^2 $, which
defines the support of $\rho $, is fixed by comparing coefficients in the
asymptotic behavior of $G$, 
\begin{equation} 
G(z) \stackrel{z\rightarrow \infty }{\longrightarrow }
\frac{1}{z} + \sum_{n=1}^{\infty } \frac{c_n }{z^{n+1 } } 
\end{equation} 
where $c_n $ is the $n$-th moment of the distribution $\rho $. One
obtains 
\begin{equation} 
Q(z) = 2a_4 z^2 + a_2 + m^2 a_4 \ \ \ \ \ \ \ m^2 = \frac{1}{3a_4 }
(\sqrt{a_2^2 + 12 a_4 } - a_2 ) 
\end{equation} 
Then one can also immediately read off
\begin{equation}
\rho (\lambda ) = -\frac{1}{\pi }  \mbox{ Im} \, G(\lambda + i\epsilon )
=\frac{1}{\pi } Q(\lambda ) \sqrt{m^2 - \lambda^{2} } 
\end{equation} 
(where $\epsilon \rightarrow 0$), and by expanding $G$ in $1/z$,
\begin{eqnarray} 
c_2 &=& \frac{1}{108 a_4^2 } \left[ -a_2^3 - 18 a_4 a_2 + (a_2^2 +
12 a_4 )^{3/2} \right] \\ 
c_4 &=& \frac{1}{216  a_4^3 } \left[ a_2^4 + 18 a_2^2 a_4 +
54 a_4^2 - a_2 (a_2^2 + 12 a_4 )^{3/2} \right] 
\end{eqnarray} 
When $a_4 =0$, these expressions simplify to
\begin{eqnarray}
\rho  (\lambda ) &=& \frac{1}{\pi  } a_2 \sqrt{(2/a_2 ) -\lambda^{2} }
\\ c_2 &=& \frac{1}{2a_2 } \\ c_4 &=& \frac{1}{2a_2^2 } 
\end{eqnarray} 
i.e. the second moment of the semicircular distribution is one fourth
of its square radius. 

In complete analogy,   one  can derive  the  two-cut  solution in  the quartic
potential. Then, $G$ has the form 
\begin{equation} 
G(z) = 2a_4 z^3 + a_2 z - q\sqrt{(z^2 -m_{+}^2 ) (z^2 -m_{-}^2 ) z^2 }
\end{equation} 
from which one again obtains by comparing coefficients in the
asymptotic expansion 
\begin{equation} 
q = 2a_4 \ \ \ \ \ \ \ m_{\pm }^{2} = -\frac{a_2 }{2a_4 } \pm
\frac{1}{\sqrt{a_4 } } 
\end{equation} 
One can immediately read off the eigenvalue distribution, and the
second moment e.g. becomes 
\begin{equation} 
c_2 = \frac{a_4 }{8}  (m_{+}^6 - m_{+}^4 m_{-}^2 - m_{+}^2 m_{-}^4 +
m_{-}^6 ) 
\end{equation}

\section{Free additive convolution of two identical distributions governed
by a quartic potential} \label{appb}  
In order to additively convolute the 
eigenvalue distributions of two matrices which are free with respect to one
another, i.e. to obtain  $\rho_{B_1 +B_2 }  $ given $\rho_{B_1 } $ and
$\rho_{B_2 } $, one should go to the corresponding R-transforms
\cite{Voi}. These are defined as follows: Find the resolvents $G_{B_1 } $ and 
$G_{B_2 } $ as defined in Appendix \ref{appa}; then the R-transforms are given
by  $R(y)=G^{-1} (y)-1/y$ (here $G^{-1} $ denotes the inverse function, $G^{-1}
(G(x))=x $). The R-transforms behave additively, i.e. 
$R_{B_1 + B_2 } = R_{B_1 } + R_{B_2 } $. In this sense
they are the free analogs of the logarithm of the Fourier
transform for probability distributions of ordinary commuting random variables. 

In the case of matrices governed by a quartic potential 
$V(x)=a_4 x^4 + a_2 x^2 $,
the resolvent has already been given explicitly in Appendix \ref{appa}; 
denoting $G(z) \equiv y$, one has that $z \equiv G^{-1} (y) $ satisfies 
\begin{equation} 
y=2a_4 z^3 +a_2 z -(2a_4 z^2 +a_2 +m^2 a_4 ) \sqrt{z^2 -m^2 }
\end{equation} 
where $m^2 $ was defined in Appendix \ref{appa}. Isolating the
square root term and squaring the resulting equation leads to a third order
equation for $z$, 
\begin{equation} 
4a_4 y z^3 - 4a_4 z^2 + 2a_2 yz - y^2 - m^2 (a_2 +m^2 a_4 )^2 = 0
\label{eqforz}
\end{equation}
It is not necessary to solve this explicitly, since the case of interest
here is the case of two identical distributions, i.e. identical R-transforms.
Then one has 
\begin{equation} 
w \equiv G^{-1}_{B_1 +B_2 } (y) = R_{B_1 + B_2 } +\frac{1}{y} =
R_{B_1 }  + R_{B_2 }   +\frac{1}{y} = G^{-1}_{B_1  } (y)  + G^{-1}_{B_2 } (y)
-\frac{1}{y} =2z-\frac{1}{y} 
\end{equation} 
implying
\begin{equation} 
z=\frac{1}{2} \left( w+\frac{1}{y} \right)
\label{zitow}
\end{equation} 
However, the equation satisfied by $z$ is known, equation
(\ref{eqforz}), which by inserting (\ref{zitow}) leads to the 
equation satisfied by $w
\equiv G^{-1}_{B_1 +B_2 } (y) $. Since one is actually interested in
$y \equiv G_{B_1 +B_2 } (w)$, one can solve directly for $y$ by rearranging 
(\ref{eqforz}) in conjunction with (\ref{zitow}) as a polynomial in $y$, 
\begin{equation} 
2y^4 - y^3 (a_4 w^3 +2a_2 w) -y^2 (a_4 w^2 + 2a_2 -2m^2 (a_2 +m^2
a_4 )^2 ) +y a_4 w +a_4 = 0 
\label{eqfory}
\end{equation} 
Rather than obtaining a cumbersome analytical solution of (\ref{eqfory}),
the relevant solution was tracked numerically. This is best done in 
small steps in along the positive
real $w$-axis starting from the known asymptotic behavior
$y=1/w +O(w^{-3}  )$ and using the positivity of the eigenvalue distribution
$\rho_{B_1 +B_2 } $. The latter is essentially given by the imaginary part 
of $y$,  
namely $\rho_{B_1 +B_2 } (\lambda ) = -\mbox{Im} \, G_{B_1 +B_2 } (\lambda
+i\epsilon ) /\pi $.

\end{document}